%% file: inv_ana_biofilm.tex
\documentclass{article}
 
\usepackage[
backend=biber,
style=numeric-comp,
date=year,
sorting=none, 
doi=true,
isbn=false,
eprint=false,
maxbibnames=99,
maxcitenames=2,
uniquelist=false,
giveninits=true,
]{biblatex}

\AtEveryBibitem{%
  \ifentrytype{book}{%
    \clearfield{url}%
    \clearfield{urldate}%
  }{}%
}
\AtEveryBibitem{\clearfield{pagetotal}}
\AtEveryBibitem{%
  \ifentrytype{misc}{}{%
    \clearfield{note}%
  }%
}

\usepackage{color}
\usepackage{xcolor}
\usepackage[colorlinks=true, linkcolor=blue, urlcolor=blue, citecolor=magenta]{hyperref}
\usepackage{graphicx}
\usepackage[utf8]{inputenc}
\usepackage{amsmath}
\usepackage{amsfonts}
\usepackage{amsthm}
\usepackage{authblk}
\newtheorem{remark}{Remark}
\usepackage[a4paper]{geometry}
\usepackage{caption,subfig}
\usepackage{booktabs}
\usepackage{array}

\usepackage{pgfplots}
\pgfplotsset{compat=1.15}
\usepgfplotslibrary[external]

\makeatother

\newcommand{\includetikz}[2]{\includegraphics[#1]{#2.pdf}}
\graphicspath{{./figures/}}

\newcommand*{\email}[1]{\normalsize\href{mailto:#1}{#1}\par}

\newcommand{\keywords}[1]{\textbf{Keywords: }#1}

\hypersetup
{
    pdfauthor={Harald Willmann},
    pdfsubject={Biofilms},
    pdftitle={Inverse analysis of biofilms},
    pdfkeywords={inverse anaylsis, biofilm}
}

\input{symbols.tex}

\begin{document}

\title{Inverse analysis of material parameters in coupled multi-physics biofilm models}

\author{Harald Willmann$^*$, Wolfgang A. Wall}

\affil{Institute for Computational Mechanics\\
Technical University of Munich\\
Boltzmannstr. 15, 85748 Garching b. München, Germany\\
e-mail: \email{harald.willmann@tum.de}
$^*$ corresponding author}

\maketitle

\begin{abstract}
In this article we propose an inverse analysis algorithm to find the best fit of multiple material parameters in different coupled multi-physics biofilm models. We use a nonlinear continuum mechanical approach to model biofilm deformation that occurs in flow cell experiments. The objective function is based on a simple geometrical measurement of the distance of the fluid biofilm interface between model and experiments. A Levenberg-Marquardt algorithm based on finite difference approximation is used as an optimizer. The proposed method uses a moderate to low amount of model evaluations. For a first presentation and evaluation the algorithm is applied and tested on different numerical examples based on generated numerical results and the addition of Gaussian noise. Achieved numerical results show that the proposed method serves well for different physical effects investigated and numerical approaches chosen for the model. Presented examples show the inverse analysis for multiple parameters in biofilm models including fluid-solid interaction effects, poroelasticity, heterogeneous material properties and growth.
\end{abstract}


\section{Introduction}

Microorganisms tend to live in aggregates rather than dispersed singe cells and surround themselves with a network of extracellular polymeric substances (EPS) as a survival strategy. This is called the biofilm matrix \cite{bio_flemming2010}. Amongst others it provides them with mechanical resistance against external forces acting on them through a surrounding fluid flow. There is a broad variety of biofilm occurrence. Depending on whether they are intentionally grown or unwillingly observed there are opposing interests when it comes to biofilm control in engineered systems. The prerequisite in all cases however is to have good knowledge about their behavior. In engineering good knowledge is represented by the availability of accurately parameterized models that allow the generation of reliable model predictions.

There has been a variety of efforts to estimate relevant material parameters of biofilms. The reader is referred to \cite{bio_boel2013}, \cite{bio_gloag2020} and  \cite{bio_boudarel2018} as they all give an overview of practiced testing methods for parameter estimation for different physical aspects. Therein a variety of mostly intrusive test strategies and results can be found. 

As it is still unknown how much influence the environment of the biofilm has on its mechanical properties \cite{bio_guelon2011}, an effort towards improved in situ testing methods has been made. The long term goal of the methods developed in this work is to develop a non-destructive testing protocol for biofilms to estimate material parameters that can later be used to predict biofilm behavior and develop strategies to control their appearance and growth. Optical coherence tomography (OCT) currently seems to be the prime option to scan the geometrical representation of biofilms in flow cell experiments under variable load \cite{kit_bio_wagner2017}. This has led to new insights into biofilm mechanics research \cite{kit_bio_blauert2015, kit_bio_picioreanu2018}. The advantage is, that automated growth and test protocols \cite{kit_bio_gierl2020} can be developed with this technique and therein the biofilm can be kept in the same environment for the whole test cycle.

First parameter estimation approaches are described in \cite{bio_stoodley1999}. A common scalar approach for determining stiffness and shear resistance is first presented therein. This type of analysis, if it is applicable, is quick to use and serves with estimates of the values in the relevant order of magnitude. While such approaches in the end might not be sufficient for predictions with suffient accuracy, they can favorably be used to assess good initial guesses for more advanced approaches like the presented inverse analysis algorithm. It has been shown in \cite{kit_bio_picioreanu2018} that fully resolved fluid-solid interaction (FSI) simulations can be used to determine one material parameter from modeling flow cell experiments with linear solid mechanics. This has the advantage that in general no specific shape characteristics of the biofilm are required for the parameter estimation.

The present work proposes an inverse analysis algorithm incorporating different numerical models for non destructive experiments with in situ measurement via  OCT. The intuitive approach followed in this work is to model the experimental setup as accurately as possible and then try to find the set of parameters that fits experimental data best, in present case via a Levenberg-Marquardt optimization. In order to best analyze and present the properties of the approach, we create a well defined, i.e. clean environment and hence use artificially generated results via forward numerical analysis and the addition of Gaussian noise as a first step and to particularly isolate important effects arising in this kind of analysis. The honest analysis and discussion of strengths and weaknesses of this type of inverse analysis approach is the main motivation to present this as standalone work.

The focus in this work is specifically set on flow cell experiments, like the ones presented in \cite{kit_bio_blauert2015} and  \cite{kit_bio_wagner2010}. In this type of experiments the mechanical load on the biofilm is controlled exclusively via the fluid volume inflow rate into the flow cell and for growth the composition of solutes in the fluid supply. OCT provides precise measurements of biofilm geometries, but it is limited in the image capturing speed, especially when it comes to three dimensional scans \cite{kit_bio_wagner2017}.

Experimental results show different shapes of biofilms for different flow rates through flow cells \cite{kit_bio_wagner2010}. Even under automated cultivation and measurement, a certain level of variability can be found  \cite{kit_bio_gierl2020}. The appearing nearly arbitrary shapes of biofilm surfaces pose the question in which way the model results should be compared to experimental results. The variety in biofilm appearance impedes the accessibility to simple or even scalar comparisons for parameter estimation.
When it comes to quantifying flow cell experiments, the load on the biofilm is sometimes represented by the fluid shear stress next to the wall of the empty flow cell. It is known \cite{kit_bio_picioreanu2018,bio_stoodley2002} and quite obvious that this shear stress is not fully representative for the load on the biofilm, as the analyzed biofilm patches in published analyses take up a significant portion of the height of the flow channel. The flow rate itself is a more representative quantity. It can also be represented by average inflow velocity or Reynolds number in the fluid flow through the empty channel. Because of the irregular shapes of biofilms, an interplay of forces and local effects in the fluid flow is affecting biofilm deformation in flow cells significantly. This is addressed in this work by full geometrical resolution of flow cell experiments and a consideration of fluid-biofilm interaction at the interface.

In order to be self contained, this article provides a very brief introduction to physical models  applicable to different aspects of biofilm mechanics and a brief summary on numerical approximations in \autoref{sec:model}. In the next step a method for comparing different surfaces is presented in \autoref{sec:measure} and the optimization used to minimize the deviation of model predictions and experimental results is depicted in \autoref{sec:lmopt}. The presented algorithm is tested on different cases and the results are discussed in \autoref{sec:examples}. General remarks and the interpretation of the informative value of results achieved with the presented algorithm are discussed in \autoref{sec:discussion}. Overall conclusions are drawn in \autoref{sec:conclusion}.

\section{Biofilm modeling}
\label{sec:model}

Properties and performance of inverse analysis approaches depend strongly on the model used to represent the experiments that are used to drive the inverse analysis. Results of an inverse analysis can never be better than the physical model incorporated. In the given setting of inverse analysis the numerical model of the experiment is referred to as the forward model. The forward model must include all physical aspects that are relevant. In the same sense all the aspects a model covers should be represented in the experimental data used.
The choice of a suitable forward model therefore depends on the type of data used and the type of application for parameters to be determined. The best case is that the model used for modeling the experiment in an inverse analysis is the same as the one, the parameters are aimed to be used with to make valid predictions for biofilm behavior in the future.

In the following we briefly sketch models and methods that we have developed in the past and that are used in this work.
Modeling approaches to fluid-solid interaction with scalar transport have been proposed in \cite{lnm_bio_yoshihara2014} and have been extended to include biofilm growth in \cite{lnm_bio_coroneo2014}.
Porous properties of biofilms and porous flow through them are known for a long time \cite{bio_deBeer1994}. For solving fluid-poroelasticity interaction (FPI) a novel method was developed in \cite{lnm_poro_ager2019a} and applied to a kind of finger shaped biofilm example therein.

The inverse analysis algorithm proposed in this paper is obviously not limited to the kind of models presented here. It can also be applied for other models, like one that describes damage in the sense of detachment \cite{bio_boel2009} or viscoelastic behavior \cite{bio_klapper2002}.

\subsection{Physical models}

The analyzed type of fluid-biofilm interaction includes significant deformations and potentially rotations of subdomains. Therefore the application of nonlinear kinematics is essential.
In addition the observations made include a variety of effects of biofilm physics and the according equations for fluid-poroelasticity interaction, scalar transport of potential nutrients and growth are summarized in the following subsections. For the sake of brevity the presentation is limited to the strong field equations. The respective boundary conditions are implicitly assumed to be well defined. Details can be found in the referenced specific literature.

\subsubsection{Fluid field}

For the fluid field the general Navier-Stokes equations for incompressible flow of a Newtonian fluid are appropriate. In order to allow for deforming domains, that are essential for fluid-solid interaction, they are given in arbitrary Lagrangean-Eulerian (ALE) formulation as
\begin{subequations}
\begin{align}
\dens^\fluidletter\partialdt{\velf}+\densf(\alevel \cdot \grad) \velf -2 \dynvisf \div \tns{\epsilon} (\velf)+\grad \presf &= \densf \bodyforce^\fluidletter&& \text{in } \domain^\fluidletter\times(0,T)\\
\div \velf &= 0 &&\text{in } \domain^\fluidletter\times(0,T).
\end{align}
\end{subequations}
These equations relate fluid velocity $ \velf $, ALE convective velocity $\alevel $, fluid pressure $ \presf $, fluid density $ \densf $, fluid dynamic viscosity $ \dynvisf $ and a body force $ \bodyforce^\fluidletter $ in the fluid domain $ \domain^\fluidletter\times(0,T) $.
Herein the strain rate tensor $\eastrain(\velf)$ is a short expression for
\begin{equation}
\eastrain(\velf)= \frac{1}{2}\left( \grad \velf + \transpose{ \left( \grad \velf \right)}\right).
\end{equation}
The ALE formulation is one popular approach to model fluid-solid interactions with a moving mesh to ensure the continuity between fluid and solid in case of a moving interface and therein the ALE convective velocity $\alevel $ is the fluid velocity relative to the moving mesh.

\subsubsection{Solid field}

For modeling the nonlinear behavior of solids also allowing for large deformations, the general balance of momentum in reference configuration
\begin{equation}
\initial{\dens}^\structletter \ddtq{\disps} =\divref (\defgrad \cdot \secpk) +\initial{\dens}^\structletter \initial{\bodyforce}^\structletter \hspace{4mm}\text{in } \initial{\domain}^\structletter\times(0,T)
\end{equation}
applies. It relates the solid density in reference configuration $ \initial{\dens}^\structletter$, solid displacement $ \disps$, deformation gradient $\defgrad $, second Piola-Kirchhoff stress tensor $\secpk$ and the body force in reference configuration $ \initial{\bodyforce}^\structletter $ in the solid domain $\initial{\domain}^\structletter\times(0,T) $.

\subsubsection{Fluid-solid interface}

On the fluid-solid interface $\interface^{\fluidletter, \structletter} \times (0,T) $ balance of tractions $\tns{h}^\structletter_\interface $ and the equality of velocities, stemming from a no slip condition between fluid and solid, need to hold.
\begin{subequations}
\begin{align}
\tns{h}^\structletter_\interface &=-\tns{h}^\fluidletter_\interface &&\text{on }\interface^{\fluidletter, \structletter} \times (0,T)\\
\partialdt{\disps}&=\velf_\interface &&\text{on }\interface^{\fluidletter, \structletter} \times (0,T)
\end{align}
\end{subequations}

\subsubsection{Scalar transport}

A scalar transport model in the coupled fluid-solid model is used for nutrient distribution during the flow cell experiment. The solution of this type of systems has been presented in \cite{lnm_bio_yoshihara2014} and \cite{lnm_bio_coroneo2014}.
The scalar transport equation is stated for the fluid \eqref{eq:scatrafluid} and the solid \eqref{eq:scatrasolid} domain.
\begin{equation}
\ddt{\conc^\fluidletter} + \alevel \cdot \grad\conc^\fluidletter - \div \left( \diffusivity^\fluidletter \grad \conc^\fluidletter\right) = 0 \hspace{4mm}\text{in } \domain^\fluidletter\times(0,T)
\label{eq:scatrafluid}
\end{equation}
\begin{equation}
\partialdt{\conc^\structletter} +\conc^\structletter\left(\div\disps\right) - \div \left( \diffusivity^\structletter \grad \conc^\structletter \right) + \react^\structletter = 0 \hspace{4mm}\text{in } \domain^\structletter \times(0,T)
\label{eq:scatrasolid}
\end{equation}
For the domains the concentration of the solute is described as $ \conc^\phasek$ in fluid and solid domains. The respective diffusion coefficients are described as $\diffusivity^\phasek$, with $\phasek \in {\fluidletter,\structletter} $ being the index for quantities in one of the phases.
$\react^\structletter$ is the reaction rate. In this work a Monod kinetic relates to nutrient consumption of the biofilm which therefore only appears in the solid domain.
\begin{equation}
\react^{\structletter, \mathrm{M}}= K_1^\mathrm{R} \frac{\conc^\structletter}{K_2^\mathrm{R} + \conc^\structletter}
\label{eq:monod}
\end{equation}
This kinetic includes two coefficients, which are the reaction rate $K_1^\mathrm{R}$ and the half saturation $K_2^\mathrm{R}$.
The negative normal flux on the interface is computed as
\begin{equation}
\flux^\phasek_\interface = \diffusivity^\phasek \grad \conc^\phasek \cdot \normal^\phasek_\interface
\end{equation}
for phases $\phasek$ with the respective unit normal $ \normal^\phasek_\interface$.
On the fluid-solid interface the concentrations and fluxes of the phases $\phasek$ must be equal.
\begin{subequations}
\begin{align}
\conc^\structletter = \conc^\fluidletter &&\text{on }\interface^{\fluidletter, \structletter} \times (0,T)\\
\flux^\structletter = \flux^\fluidletter &&\text{on }\interface^{\fluidletter, \structletter} \times (0,T)\label{eq:scalarinterfaceluxes}
\end{align}
\end{subequations}
In this formulation \eqref{eq:scalarinterfaceluxes} the fluxes on the interface must be related to the same normal direction on the interface. The scalar transport is only one way coupled to the fluid-solid interaction as the deformation of the solid and the fluid velocity will influence the concentration solution, but the concentration does not really influence solid or fluid field solutions.

\subsubsection{Surface growth}

We introduce growth or erosion according to \cite{lnm_bio_coroneo2014}. It is assumed that growth or erosion is localized on the biofilm surface and influenced by nutrient flux and surface tractions. 
\begin{equation}
\disps_\growthindex = \Delta t^\growthindex \left(K_1^\growthindex \flux^\structletter -  K_2^\growthindex \abs{ \left( \cauchy^\structletter \cdot \normal^\structletter \right) \cdot \normal^\structletter} - K_3^\growthindex \abs{ \sum_{i=1}^2\left( \cauchy^\structletter \cdot \normal^\structletter \right) \cdot \tangent_i^\structletter} \right) \normal^\structletter
\label{eq:3compgrowthlaw}
\end{equation}
With $ \normal^\structletter$ being the outward pointing normal on the biofilm surface and $\tangent_i^\structletter $ two respective orthogonal unit tangents. This is a relatively simple phenomenological model wherein stress inhibits growth or erodes the biofilm \cite{lnm_bio_coroneo2014} and the nutrient flux into the biofilm domain is the cause for growth. The normal flux contributes to growth with the factor $K_1^\growthindex $ and the erosion is determined with the factors $ K_2^\growthindex$ due to the normal stress component and  $ K_3^\growthindex$ due to the tangential stress components.  $\Delta t^\growthindex$ describes the timespan the biofilm is exposed to given growth conditions and $\disps_\growthindex $ the resulting displacements of the surface because of growth.

\subsubsection{Poroelasticity field}

The poroelasticity model, that is used, relies on the assumption of a homogenized mixture between a fluid phase with Darcy flow and a solid structure. Porosity $\porosity$ acts as volume ratio of void that is filled with a fluid phase. This is well described in \cite{poro_coussy2003} and a fully coupled numerical model for the field equations was developed in \cite{lnm_poro_vuong2015}.
The coupled system of equations for the poroelastic mixture is derived as
\begin{subequations}
\begin{align}
\evalat{\partialdt{\porosity}}{\poroxref}+ \porosity \div \partialdt{\porodisp} + \div\left[\porosity\left(\porovel - \partialdt \porodisp \right) \right] &= 0 &&\text{in }\domain^\poroletter \times (0,T) \label{eq:porofluid}\\
\dens^\porofluidletter \evalat{\partialdt{\porovel}}{\poroxref} - \dens^\porofluidletter \partialdt{\porodisp} \cdot \grad \porovel + \grad \poropres - \dens^\porofluidletter {\bodyforce}^\porofluidletter &\ldots&&\nonumber\\
\ldots + \dynvis^\porofluidletter \porosity \inverse\poropermeab \cdot \left(\porovel - \partialdt{\porodisp} \right) &=\tns0 &&\text{in }\domain^\poroletter \times (0,T) \\
\tilde{\initial{\dens}}^\porostructletter \partialdtdt\porodisp - \divref \left( \defgrad \cdot \secpk^\poroletter\right) - \tilde{\initial{\dens}}^\porostructletter \initial{\bodyforce}^\poroletter - \defjac \porosity \invtrans\defgrad \cdot \initial\grad \poropres &\ldots&&\nonumber\\
\ldots- \dynvis^\porofluidletter \defjac \porosity^2 \inverse\poropermeab \cdot \left(\porovel - \partialdt{\porodisp} \right) &=\tns0 &&\text{in }\initial{\domain}^\poroletter \times (0,T). 
\label{eq:porostruct}
\end{align}
\label{eq:poroequations}
\end{subequations}
In \eqref{eq:poroequations} the indices $(\bullet)^\porostructletter $ for the porous structure phase, also called skeleton and $ (\bullet)^\porofluidletter$ for the fluid phase are used. $ J = \det{\defgrad} $ is generally known as the Jacobian determinant being the determinant of the deformation gradient $\defgrad$. With the presented formulation the porosity needs an initial value for each material point and then changes according to the skeleton displacement field of the fully coupled model during deformation.
$\tilde{\initial{\dens}}^\porostructletter = (1-\initial{\porosity})\initial{\dens}^\porostructletter  $ represents the macroscopic averaged initial density of the skeleton and $ \poropermeab = \inverse{(\defjac)} \defgrad \cdot \matpermeabp \cdot \transpose{\left(\defgrad\right)}$ is the spatial permeability computed from the permeability $\matpermeabp$ in reference configuration, which is determined with the Kozeny-Carman formula.
\begin{equation}
\matpermeabp = \identity \matpermeabpscalar \frac{1 - \initial\porosity^2}{\initial\porosity^3} \frac{(\defjac \porosity)^3}{1 - (\defjac \porosity)^2}
\end{equation}
The porous composite strain energy function is
\begin{equation}
\strainenergyp(\glstrain, \defjac(1-\porosity)) = \strainenergy^{\poroletter\mathrm{,skel}}(\glstrain) + \strainenergy^{\poroletter\mathrm{,vol}}(\defjac(1-\porosity)) + \strainenergy^{\poroletter\mathrm{,pen}}(\glstrain, \defjac(1-\porosity)).
\end{equation}
Therein $\strainenergy^{\poroletter\mathrm{,skel}}$ is the contribution from the skeleton, $\strainenergy^{\poroletter\mathrm{,vol}}$ accounts for the volume change due to changing fluid pressure and the penalty part $ \strainenergy^{\poroletter\mathrm{,pen}}$ guarantees positive porosity.
This formulation allows usage of any hyperelastic material model for the skeleton part. We choose the scaling parameters as penalty parameter $=1.0 $ and bulk modulus $=100 $ accoring to \cite{lnm_poro_vuong2015} for the presented example.

\subsubsection{Fluid-poroelastic solid interaction}

A consistent approach to tackle the interaction of a Newtonian fluid and a poroelastic solid is presented in \cite{lnm_poro_ager2019a}. The method presented therein is directly applied in this work. On the interface
\begin{subequations}
\begin{align}
\cauchyf \cdot \normal^\fluidletter - \cauchy^\poroletter \cdot \normal^\fluidletter &= \tns 0 &&&\text{on }\interface^{\fluidletter, \poroletter} \times (0,T)\label{eq:porotracbalance}\\
\normal^\fluidletter \cdot \cauchyf \cdot \normal^\fluidletter + \poropres &= 0 &&&\text{on }\interface^{\fluidletter, \poroletter} \times (0,T) \label{eq:poropresbalance}\\
\left[\velf - \partialdt{\porodisp} -\porosity \left( \porovel - \partialdt \porodisp \right) \right]  \cdot \normal^\fluidletter &= 0 & &&\text{on }\interface^{\fluidletter, \poroletter} \times (0,T)\label{eq:poronormalflow}\\
\left[\velf - \partialdt{\porodisp} -\beta_\beaversjoseph \porosity \left( \porovel - \partialdt \porodisp \right) +  \fpipermeab \normal^\fluidletter \cdot \cauchyf \right]  \cdot \tangent^\fluidletter_i &= 0 &i=1,2 &&\text{on }\interface^{\fluidletter, \poroletter} \times (0,T) \label{eq:BJtangvel}
\end{align}
\end{subequations}
must hold. The conditions describe a balance of tractions between the porelastic mixture and the pure fluid \eqref{eq:porotracbalance}, equality of fluid pressure in the poroelastic and fluid domain \eqref{eq:poropresbalance}, the continuity equation for the normal fluid flow \eqref{eq:poronormalflow} and the so called Beavers-Joseph conditions \cite{poro_beavers1967} for the coupling of the tangential components in directions $\tangent^\fluidletter_i$ of the fluid velocities \eqref{eq:BJtangvel}. 
Therein the interface permeability $\fpipermeab$
\begin{equation}
\fpipermeab = \inverse{\left(\alpha_\beaversjoseph \dynvisf \sqrt{3} \right)} \sqrt{\trace{\poropermeab}}
\end{equation}
is used. The Beavers-Joseph constants $\alpha_\beaversjoseph, \beta_\beaversjoseph$ regulate this tangential velocity dependency in \eqref{eq:BJtangvel}. They are both chosen to $\alpha_\beaversjoseph = \beta_\beaversjoseph =1 $ in the presented example.

\subsection{Numerical approximation}

For all numerical models a nonlinear finite element method (FEM) based approach is used. For time integration a one-step-theta approach is used.
For the pure FSI examples we use a monolithic arbitrary Langrangean-Eulerian (ALE) approach just as in \cite{lnm_bio_yoshihara2014} and \cite{lnm_bio_coroneo2014}. For the mesh deformation the ALE field can be treated as a quasi-elastostatic pseudo solid.
Monolithic methods are preferable for FSI problems in many biological applications as they might contain fields with similar density with soft solids, represented here by low Young's modulus \cite{lnm_fsi_kuettler2010}. 

ALE Methods can be problematic when it comes to a change in the topology or large mesh displacements.
An alternative approach to overcome difficulties associated with these cases are fixed grid methods. For the fluid-poroelasticity interaction examples presented, a CutFEM based approach \cite{lnm_poro_ager2019a} is used. For FSI problems an approach based on the so called CutFEM has been developed in recent years \cite{lnm_fsi_schott2019}. It is capable to solve FSI problems with a fixed fluid grid. The idea is to cut out the parts of the fluid mesh that are covered by the solid and solve the fluid field on the remaining discretization including partially cut elements at the interface. To sustain a proper fluid mesh with uncut elements in the interface vicinity also a hybrid method of ALE and CutFEM was developed in \cite{lnm_fsi_schott2019a}.
CutFEM based FSI approaches enable the treatment of cases when it comes to a change in topology like for partial or full detachment or on the other side self contact \cite{lnm_fsi_ager2020,lnm_poro_ager2019}.

For the growth algorithm there is the need for multiple time scales as the FSI dynamics acts in the range of seconds and the growth processes take place in the range of days. For this multi-scale approach in time a quasi steady or periodic state for the fluid-solid-scalar interaction is reached with smaller time steps. Based on the FSI and scalar transport solution the nutrient flux and interface tractions are evaluated and surface growth is applied with the much longer growth time step. This procedure is then advanced until the full growth time period is reached. After the growth step an ALE relaxation of the mesh is necessary for the domains on both sides of the interface fluid and biofilm, to distribute the displacements smoothly on the whole domains and thereby reduce mesh distortion \cite{lnm_bio_coroneo2014}.

\section{Surface distance measure}
\label{sec:measure}

We set inverse analysis as a special optimization problem and in the application of optimizers the quantity that is minimized is called objective function \cite{inv_nocedalwright2006}. In inverse analysis the objective function somehow describes the difference between experimental observation and forward model output. In this context it turns out that the measure that is used to quantify this difference is a key question in inverse analysis. The result of an inverse analysis always depends on the approach used for this measure. Hence, detailed information about the measurement approach need to be combined with the achieved results for presentation and interpretation. This also makes it obvious that the selection or design of a suitable measure is crucial and must be well considered. One important contribution of this work is to propose a simple geometric measurement for the objective function in this kind of experimental settings for biofilm parameter estimation problems, that is suitable for any optimizer.

\begin{figure}[htbp]
\centering
\includegraphics[width=0.6\textwidth]{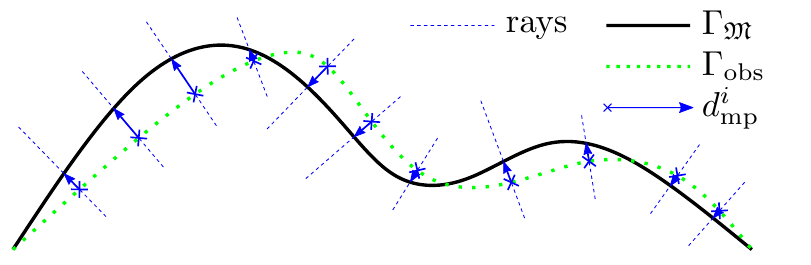}
\caption{Prototype sketch for pointwise distances $\signeddist_\mpindex^i$ from measurement points on the deformed interface $ \interface_\obsindex $, observed from experiment, to $\interface_\forwardmodelletter$ resulting from forward model evaluation, in given measurement directions depicted as rays.}
\label{fig:shape}
\end{figure}

As OCT measurements of experiments only contain information if a point in space is likely covered by biofilm or not, there is no pointwise displacement information available. Comparable pointwise displacements from the observed experiment would need to be somehow computed first. But in order to do so, some additional assumptions would need to be introduced, which in turn spoil or bias the outcome. In addition those assumptions - being more or less physical - might even substantially complicate the inverse problem or they might point to "unphysical" scenarios. Because of this we argue that it is not the best idea to compare the forward model evaluations to a field of selected point displacements, which are themselves the result of a postprocessing operation, but rather refer to some primary information, which, in the case of flow cell experiments and OCT, is the surface shape.
Evaluating this information is performed as depicted in \autoref{fig:shape}.

Given an observed result of an OCT measurement the first step is to determine a representation of the fluid-biofilm interface $ \interface_\obsindex$, by some sort of image segmentation. Image segmentation is a topic on its own and not the focus of this work. For the purpose of this paper, it is enough to assume that the data is already suitably segmented, which can also mean a segmentation done by hand. On the observed interface $ \interface_\obsindex $ the analyst is to choose significant points, meaning points where the interface underwent significant displacements during the experiment. These measurement points are depicted as crosses in \autoref{fig:shape}. The distribution and number of measurement points is up to the choice of the analyst as it depends on many aspects. The number should at least be greater than the number of parameters, that are to be determined in the inverse analysis for the optimizer to work well. The actual choice of measurement points should in our impression be made towards regions, where the validity of all model assumptions is trusted the most. That means it should favor points away from potentially uncertain boundary conditions and potentially uncertain flow conditions in the channel and towards regions, where the spatial resolution of OCT and the quality of the captured images is trusted the best. Secondly, given the forward model and the parameters analyzed, the measurement points should be chosen in a way that all the different parameters can show significant effect in the resulting distances.

For every point selected, an individual search direction for the intersection with the result for the displaced fluid-biofilm interface from the forward model evaluations $\interface_\forwardmodelletter $ must be decided. These are depicted as rays in \autoref{fig:shape}. A general recommendation is the normal direction. Nevertheless again the quality of the image decides if a confident guess for that normal can be made. Given the fact, that OCT scans are generated from above the experiment, the vertical direction is another reasonable choice.

With both the measurement points and the associated directions at hand the thereby defined rays must be intersected with the deformed interface resulting from forward model evaluations $ \interface_\forwardmodelletter$. As the resulting distance for every measurement point  $\signeddist_\mpindex^i$ is based solely on a geometrical measure, it has no inherently conclusive sign. Therefore it is defined positive if the intersection point is on the side of the biofilm with respect to $ \interface_\obsindex $ and negative if it lies towards the outside. In the case of multiple intersection points the lowest resulting distance is used.

The presented choice of comparative measure for the interfaces including both measurement point location and measurement direction circumvents known drawbacks of the closest point projection described in \cite{inv_imperiale_2013}. With the proposed method the distance measurements are uniquely defined as the search direction is predefined. This can be very useful as, if two or more candidates for the closest point on $\interface_\forwardmodelletter$ to a measurement point exist, a gradient based optimization with a finite difference approximation as the one presented, can be heavily deteriorated. The capturing of irrelevant shape characteristics must be prevented by a good choice of measurements points by the user.

Overall the presented measurement method is considered rather hands-on, as the observed interface location must only be determined for the measurement points. For our type of problems, this is also a clear advantage compared to the usage of global surface comparisons as for example the ones presented in \cite{inv_vaillant2005} and \cite{lnm_inv_kehl2016}. For global approaches for surface comparison a full representation of the surface must be available and therefore must be constructed from the data. A global measurement approach also poses higher demands on the image segmentation than the presented method. Nevertheless in the case of optimal data acquisition and the assumption that the experiment is modeled optimally, the presented measurement is also fully automatable using factual normals e.g. for every triangle of a triangulation of the observed deformed biofilm surface $ \interface_\obsindex$.

\section{Levenberg-Marquardt optimization}
\label{sec:lmopt}

For the minimization of the objective function defined by a suitable comparison of observed experiments and a forward model we use a Levenberg-Marqaurdt approach for optimization. Levenberg-Marquardt optimizers go back to the works of Levenberg \cite{inv_levenberg1944} and Marquardt \cite{inv_marquardt1963}. A good overview of the actual algorithm is shown in \cite{inv_more1978}. A Levenberg-Marquardt optimizer is in general a deterministic method. As a gradient based method it represents a local optimizer and is applicable for inverse analysis if the dimension of the inverse problem is low enough and the initial guess is good enough. In the selected numerical examples we will shed some light on the applicability for different numbers of parameters and initial guesses for our target applications. In the past we have already succesfully applied such algorithms for identification of constitutive laws and parameters of hyper- and visco-elastic biomechanical problems (see e.g. for problems with single type of experiments \cite{lnm_inv_rausch2011,lnm_inv_bel2014} and also the combination of different experiments on the same specimen \cite{lnm_inv_birzle_2019a}).

In order to have a rather self contained paper we will briefly sketch the algorithm in the following. The Levenberg-Marquardt method is used to minimize a least squares objective function 
\begin{equation}
f(x)=\frac{1}{2} \sum_{j=1}^\numres r_j^2(\tns{x})=\frac{1}{2} \sum_{j=1}^\numres \left(\model{\paramvec}_j-\left(y_\obsindex\right)_j\right)^2
\end{equation}
with forward model results $\model{\paramvec}_j $ at potentially different times for $\numparam $ unknown parameters in the parameter vector $\paramvec $ and $\numres $ observed experimental measurements $\left(y_\obsindex\right)_j$.
The algorithm uses the regularization parameter $\mu$ and is started from an initial guess $\initialtop{\paramvec}, \initialtop{\mu}$.
The core algorithm is to iterate the update rule for the parameter vector
\begin{equation}
\paramvec^{k+1}=\paramvec^k+\Delta\paramvec^{k+1}
\end{equation} 
with the parameter step computed by
\begin{equation}
\Delta\paramvec^{k+1}= -\inverse{\left( \transpose\jacmat \cdot \jacmat + \mu \diag{\transpose\jacmat \cdot \jacmat }\right)}\transpose\jacmat\lmres
\end{equation}
until predefined convergence criteria are met. The iteration index $k$ is omitted in $\jacmat^k$, $ \mu^k$ and $\lmres^k $ from here, as it is clear that terms should be computed exclusively at current step $k$.
We propose that the vector of punctual distances $\signeddist_\mpindex^i$, potentially also collected over different time steps, between the forward model result surface and the observed experimental surface measured in the way presented in \autoref{sec:measure} should be directly used as residuals and arranged in the residual vector $\lmres$ of length $\numres$.

\begin{equation}
\lmres=
\begin{bmatrix}
\signeddist_\mpindex^1\\
\vdots\\
\signeddist_\mpindex^\numres\\
\end{bmatrix}
\end{equation}

The Levenberg-Marquardt method makes good use of the vector shape of the residual for finding the parameters for the next step. 
The algorithm uses the partial derivatives of the residual components with respect to the parameters as the Jacobian

\begin{equation}
\jacmat=
\begin{bmatrix}
\partiald{r_1}{\paramletter_1}& \hdots & \partiald{r_1}{\paramletter_\numparam}\\
\vdots & \ddots &\vdots \\
\partiald{r_\numres}{\paramletter_1}& \hdots & \partiald{r_\numres}{\paramletter_\numparam}\\
\end{bmatrix}.
\end{equation}
In absence of actual gradient information, the Jacobian is approximated by finite differences. For that, $\numparam+1$ simulations per iteration are necessary to be computed.
One model evaluation is conducted with the current parameter set
$
\paramvec \hspace{0.2cm}(=\paramvec^k)
$
. $\numparam$ further model evaluations are computed with the $i^\mathrm{th}$ parameter perturbed to
\begin{equation}
\tilde{\paramletter}_i=\paramletter_i+\alpha+\beta \paramletter_i \hspace{0.5cm}\Rightarrow \delta \paramletter_i = \tilde{\paramletter}_i - \paramletter_i =\alpha+\beta \paramletter_i.
\label{eq:perturbation}
\end{equation}
Resulting $\numparam$ perturbations of the parameter are written as vectors $\paramvec_i$.
The approximations of the partial derivatives
\begin{equation}
\partiald{r_j}{\paramletter_i}\approx
\findiffs{r_j}{\paramletter_i}=
\dfrac{r_j(\paramvec_i)-r_j(\paramvec)}{\alpha + \beta \paramletter_i}
\end{equation}
are then arranged into the approximation of the Jacobian
\begin{equation}
\jacmat\approx
\begin{bmatrix}
\findiffs{r_1}{\paramletter_1}& \hdots & \findiffs{r_1}{\paramletter_\numparam}\\
\vdots & \ddots &\vdots \\
\findiffs{r_\numres}{\paramletter_1}& \hdots & \findiffs{r_\numres}{\paramletter_\numparam}\\
\end{bmatrix}
=
\begin{bmatrix}
\dfrac{r_1(\paramvec_1)-r_1(\paramvec)}{\alpha + \beta \paramletter_1}
& \hdots 
& \dfrac{r_1(\paramvec_\numparam)-r_1(\paramvec)}{\alpha + \beta \paramletter_\numparam}\\
\vdots & \ddots &\vdots \\
\dfrac{r_\numres(\paramvec_1)-r_\numres(\paramvec)}{\alpha + \beta \paramletter_1}
& \hdots 
& \dfrac{r_\numres(\paramvec_\numparam)-r_\numres(\paramvec)}{\alpha + \beta \paramletter_\numparam}\\
\end{bmatrix}.
\end{equation}
To assess how much the current step has improved the result, the gradient based error $ \graderr^{k}$
\begin{equation}
\left(\norm{\transpose\jacmat\lmres}{2}\right)^{k}:= \graderr^{k}
\label{eq:graderr}
\end{equation}
can be used. In order to simply evaluate how close the current model solution is to the experiment, the residual error $\reserr^{k}$ can be computed as
\begin{equation}
\left(\frac{\norm{\lmres}{2}}{\sqrt{\numres}}\right)^{k}:= \reserr^{k}.
\end{equation}
The regularization parameter is updated according to
\begin{equation}
\mu^{k+1}= \mu^k\frac{\left(\norm{\transpose\jacmat\lmres}{2}\right)^k}{\left(\norm{\transpose\jacmat\lmres}{2}\right)^{k-1}}
\end{equation}
only if the current step is closer than the previous step.
\begin{equation}
\graderr^{k}<\graderr^{k-1}, \reserr^k < \reserr^{k-1}
\end{equation}
Treating the regularization parameter in this adjusting way is the only form of adaptivity in the algorithm. In general it helps to find a suitable step size by reducing the regularization especially close to the optimum.

So far the presented Levenberg-Marquardt method shares the property with its family of optimizers to be unbounded. No information about the validity of parameters is introduced so far. Often material parameters come with a valid interval, or constraints, that need to be fulfilled. The pure algorithm presented so far is not constrained, so potentially if the Jacobian indicates further decrease of the residual into a given direction, the algorithm suggests parameters $\paramvec^{k+1}$, that cannot be used in the model. On top of the Levenberg-Marquardt algorithm we want to be able to set constraints on every parameter.
\begin{equation}
\imin{\paramvec} < \paramvec^{k} < \imax{\paramvec}
\label{eq:bounds}
\end{equation}
To achieve this property an additional check is introduced. If any suggested parameter in $\paramvec^{k+1}$ is out of bounds, the step is declined, the regularization parameter is doubled $ \mu^k = 2 \mu^k$ and a new step $\paramvec^{k+1}$ is proposed. To avoid useless model evaluations the algorithm is terminated if $\mu^k$ grows unreasonably high $\mu^k > \initialtop{\mu} \cdot 10^6 $. In that case no parameter result can be found. 

The algorithm is terminated if a certain convergence criterion is met or if a maximum number of iterations is $\mathrm{n}_\mathrm{max} $ reached
\begin{equation}
\lmgradcrit>\graderr^k
\end{equation}
\begin{equation}
\lmrescrit>\reserr^k
\end{equation}
\begin{equation}
k>\mathrm{n}_\mathrm{max}.
\end{equation}
For real experimental data, comparison to $\epsilon_{\mathrm{grad}}$ is the better suited criterion, because the residual error in the data is unknown. This way iterations are stopped, when the gradient does not indicate significant decent in the residual measure. As $\graderr$ has by definition \eqref{eq:graderr} no unique physical unit as it depends on different parameters in $\paramvec$, physical units are omitted for this quantity.  Overall the algorithm is deterministic and a local optimizer. So, if more than one local optimum exists within or outside of the bounds, there is no guarantee to find a global optimum even for a continuous and bounded problem.

\section{Numerical examples}
\label{sec:examples}

In the following numerical examples we want to show that given inverse problems in biofilm physics are well solvable with the presented measure for the similarity of surfaces from \autoref{sec:measure} and the Levenberg-Marquardt approach presented in \autoref{sec:lmopt}. The general setup is representing a prototype flow cell experiment, wherein a solid representing the biofilm is exposed to a certain volume flow rate from the left and is therefore deformed towards the right. As already stated above, the performance of an inverse analysis depends on a number of things beside the method itself, like the question how well the numerical model represents reality in the experimental setup. Hence in order to get a better impression of the quality of a specific method for a certain type of application, it is advantageous to test such methods on "clean data" first. A common approach to generate such data is to use the numerical model in a forward analysis with some chosen parameters and potentially add some noise to the results, in order to generate artificial experimental or measurement data to be used in the following inverse analyses.

The numerical examples were computed using the referenced methods implemented in the inhouse multi-physics C++ code BACI \cite{BACI} and a tailored python framework QUEENS \cite{QUEENS}. The presented inverse analysis algorithm was newly implemented in QUEENS during this work. QUEENS is used to run and manage forward model evaluation and conduct the inverse analysis. The intersections of the rays representing the measurement directions and the mesh based forward model results have been found with the python package for vtk \cite{vtk}.

Although the methods are implemented and fully capable of handling three dimensional geometries, for the sake of presentation the examples are limited to two dimensional effects in purely 2D and quasi 2D (i.e. 3D with just one layer of elements in the third dimension and according boundary conditions) examples. For the finite difference scheme \eqref{eq:perturbation} $\alpha = 10^{-5}$, $\beta = 10^{-3}$ are chosen for all examples. The fluid is modeled with $\dynvisf = 10^{-3}\sunit{Pa\, s} $ and $ \densf = 10^3\sunitfrac{kg}{m^3} $ for water. The material models for biofilms share the density of water. The flow cell experiments are modeled to last several seconds. The focus is set on the quasi static case, meaning that the biofilm is free of oscillatory or inertia effects in the observed reference results from forward model evaluation. For that, the inflow rate is applied smoothly with a cosine based function in multiple steps of an increasing volume inflow and then held constant. The inflow is assumed to be the result of laminar channel flow and therefore chosen to have a parabolic profile. This is how the inflow boundary condition can be reduced to a single quantity, the volume inflow rate.  On the right hand side boundary a horizontal outflow is enforced, to reflect, that it is no free outflow, but the channel continues downstream from the modeled region.

\subsection{Homogeneous biofilm}

As a first and most simple example the presented inverse analysis algorithm is performed for a fully homogeneous solid model of a biofilm interacting with the fluid flow. The shape of the biofilm domain in the simulated model is arbitrary and inspired by experiments shown in \cite{kit_bio_blauert2015} \cite{kit_bio_picioreanu2018}. A parabolic fluid flow profile with flow rate $100\sunitfrac{mm^2}{s} $ from the left boundary of the purely two-dimensional channel with $2\sunit{mm}\times1\sunit{mm} $ is set as a load.

For the presented FSI examples a Saint-Venant-Kirchhoff material is used like in other biofilm related works \cite{bio_boel2009, lnm_bio_taherzadeh2010}. Its behavior is governed by two parameters, namely Young's modulus $\youngs $ and the Poisson's ratio $\poissonratio$. This material is linear in the Green-Langrange strains and second Piola-Kirchhoff stresses and for cases with small deformations can be related to the classical Hooke constitutive law, which is standard in linear continuum mechanics and also used in other biofilm mechanics studies \cite{kit_bio_picioreanu2018}. The biofilm is obviously three-dimensional and it is assumed that its behavior is not changing much in the out of plane direction, which is reflected by using a plain strain assumption for the reduction to two dimensions.

\begin{figure}[h!]
\centering
\subfloat[\label{subfig:FSIveldisp}]{
\includegraphics[width=0.49\textwidth]{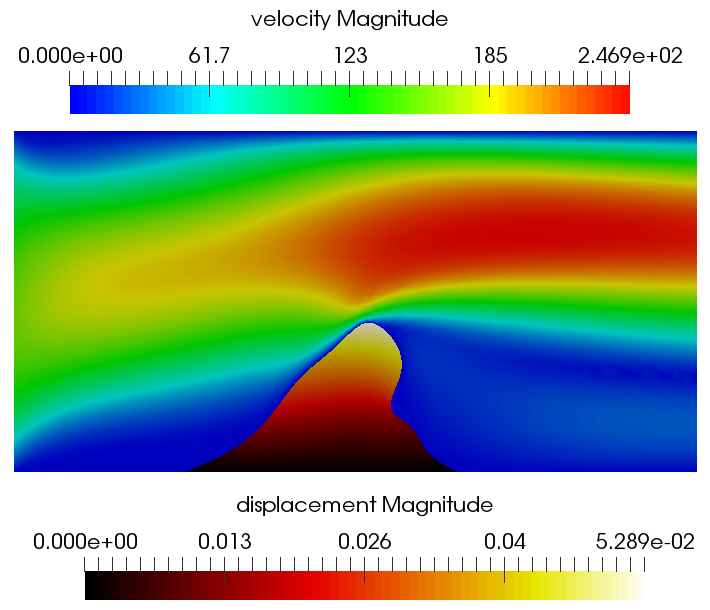}}
\hfill
\subfloat[\label{subfig:FSIprestrac}]{
\includegraphics[width=0.49\textwidth]{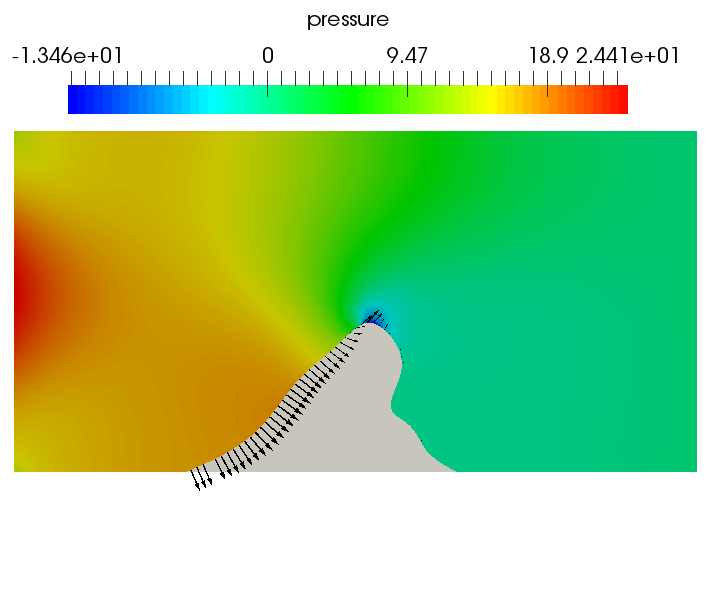}}
\caption{Field solutions of the reference simulation illustrating the deformed geometry. (a) Fluid velocity magnitude and structure displacement magnitude. (b) Fluid pressure solution and interface tractions on the biofilm as black arrows.}
\label{fig:FSIresult}
\end{figure}

\begin{figure}[h!]
\begin{minipage}[t]{0.5\textwidth}
\centering
\includegraphics[width=0.7\textwidth]{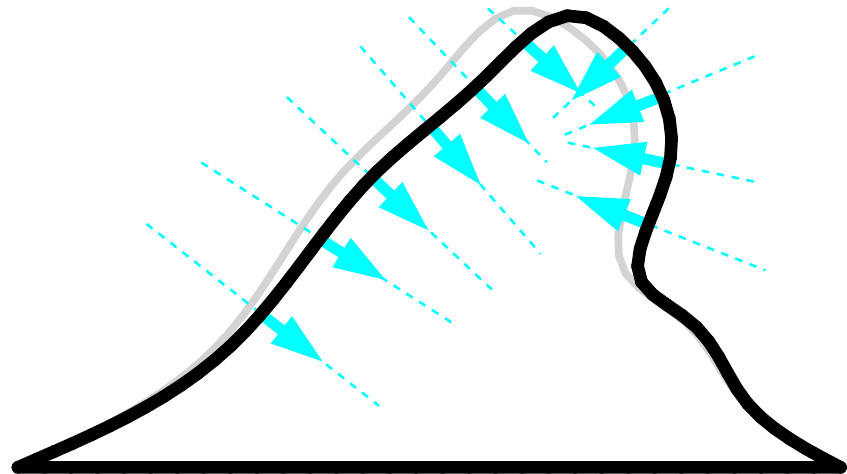}
\caption{Measurement points and directions as arrows indicating the positive directions for homogeneous FSI example on the deformed geometry of the reference result (black) and initial geometry (gray).}
\label{fig:FSIarrows}
\end{minipage}
\hfill
\begin{minipage}[b]{0.45\textwidth}
\captionsetup{type=table}
\caption{Initial guesses for different inverse analysis runs in homogeneous example.}
\centering
\input{table_initial_guesses.tex}
\label{tab:initial_guesses}
\end{minipage}
\end{figure}

First, the reference result with parameters $ \youngs = 400 \sunit{Pa}$ and $ \poissonratio = 0.3 $ in the biofilm model is computed and the field solutions shown in \autoref{fig:FSIresult} are obtained. In \autoref{subfig:FSIprestrac} additionally the interface tractions acting on the biofilm, resulting from the fluid-solid coupling, are displayed as arrows. The simplistic assumption commonly used in biofilm mechanics of a constant tangential force on the whole fluid biofilm interface would be distinctively inaccurate in this example, as it can be observed that the interface tractions are predominantly normal to the interface. The tangential component of the interface tractions varies strongly at the interface. The resulting change in biofilm shape is illustrated in \autoref{fig:FSIarrows}.

To apply the inverse analysis algorithm, pairs of significant points and the measurement directions on the deformed interface must be chosen. Those are chosen as shown in \autoref{fig:FSIarrows} on the deformed interface of the reference solution. The points where the rays (dotted lines) cut the interface in \autoref{fig:FSIarrows} represent the location of the measurement points and the arrows indicate the direction in which the distance measure is evaluated as positive for respective forward model outputs $\interface_\forwardmodelletter $. The points are evenly distributed in regions with significant displacement on the upstream and downstream side of the biofilm. Directions are chosen to be normal on the displaced geometry.

\begin{figure}[htbp]
\centering
\subfloat[\label{subfig:FSIpathne}]{\includetikz{width=0.4\textwidth}{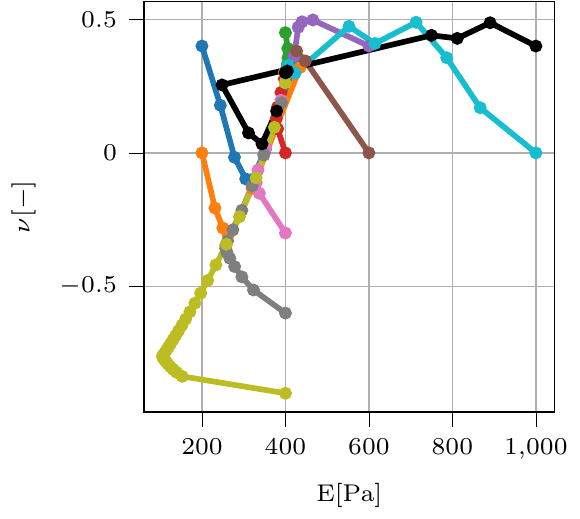}}
\hspace{1cm}
\subfloat[\label{subfig:FSIpathe-2}]{\includetikz{width=0.402\textwidth}{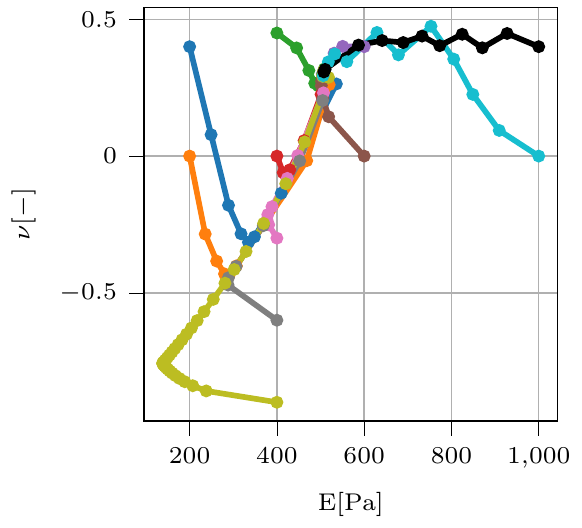}}
\caption{Path in parameters space towards optimum for data sets (a) without noise and (b) with Gaussian noise with standard deviation $\sigma=10^{-2}\sunit{mm}$ added to the measurements with one marker per iteration.}
\label{fig:FSIpath}
\end{figure}

With the given prerequisites the inverse analysis of the two material parameters Young's modulus $\youngs $ and the Poisson's ratio $\poissonratio$ is conducted with different initial guesses listed in \autoref{tab:initial_guesses}. The cases with negative Poisson's ratio are included as there is some speculation in the literature wether biofilms are so called auxetic materials. Resulting search paths in the parameter space are shown in \autoref{fig:FSIpath} with the associated colors, which are consequently used throughout this example. In \autoref{fig:FSIpath} and all following plots each marker represents one Levenberg-Marquardt iteration.

To assess the impact of noise in the data on the inverse analysis result, different scales of normally distributed (Gaussian) noise with standard deviations $\sigma$ of $ 10^{-4}\sunit{mm}$, $ 10^{-3}\sunit{mm}$ and $ 10^{-2}\sunit{mm}$ have been added to the measured points representing the experimental data. For all data sets the algorithm has been run for the same initial guesses. The results are summarized in \autoref{tab:values} for statistics over all algorithm runs with all different initial guesses for the individual noise levels and the noise free case. With increasing noise on the data the remaining residual error $\reserr$ increases. Means of the parameter results drift away from the ones used for the reference forward model evaluation and the respective standard deviations in the parameter results increase.

The actual resolution of OCT measurements is in the range of $\unit{\mu m}$ \cite{kit_bio_wagner2017, kit_bio_gierl2020}. In this simplest of the presented examples and this primitive uncertainty estimation in \autoref{tab:values} it appears that the expected accuracy for the inverse analysis with this level of noise reaches at best the first two digits of the parameter results.

\begin{table}[htbp]
\caption{Mean values and standard deviations (std) over all runs with different initial guesses per data set with added Gaussian noise with standard deviation $\sigma$.}
\centering
\input{table_res_E_nue.tex}
\label{tab:values}
\end{table}

\begin{figure}[htbp]
\centering
\subfloat[\label{subfig:FSIreserr}]{
\includetikz{width=0.3\textwidth}{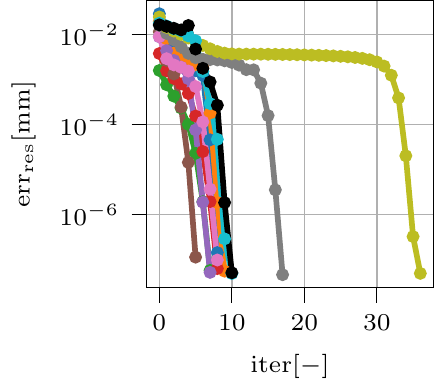}}
\hfill
\subfloat[\label{subfig:FSIgraderr}]{
\includetikz{width=0.31\textwidth}{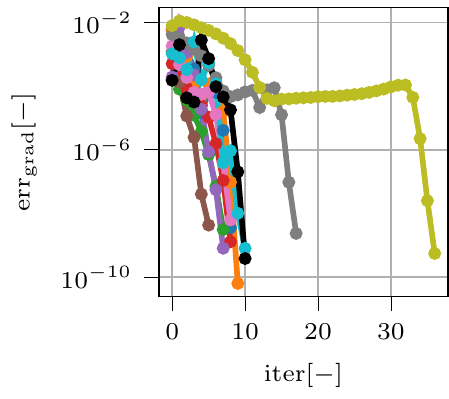}}
\hfill
\subfloat[\label{subfig:FSImu}]{
\includetikz{width=0.303\textwidth}{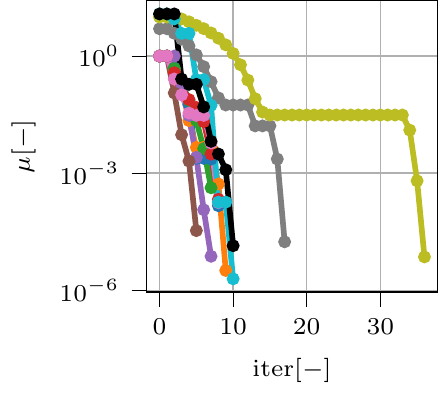}}
\caption{Development of (a) residual error $\reserr$, (b) gradient based error $\graderr$ and (c) regularization parameter $\mu$ for inverse analysis runs with different initial guesses over iterations.}
\label{fig:FSIerrors}
\end{figure}

\begin{figure}[htbp]
\centering
\subfloat[\label{subfig:FSIreserre-4}]{
\includetikz{width=0.3\textwidth}{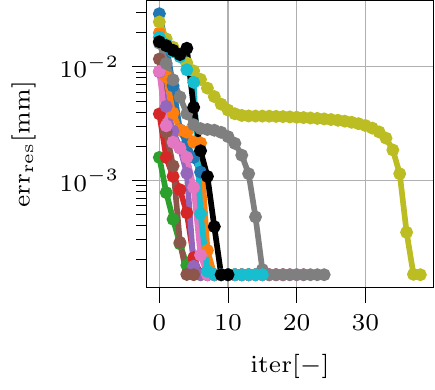}}
\hfill
\subfloat[\label{subfig:FSIgraderre-4}]{
\includetikz{width=0.305\textwidth}{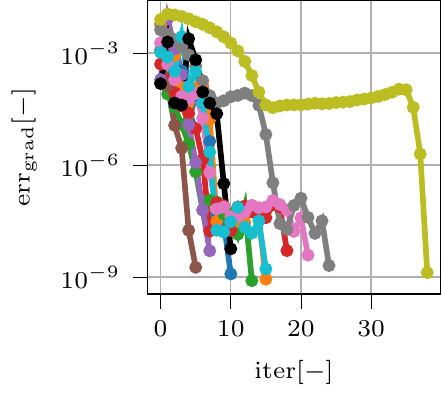}}
\hfill
\subfloat[\label{subfig:FSImue-4}]{
\includetikz{width=0.305\textwidth}{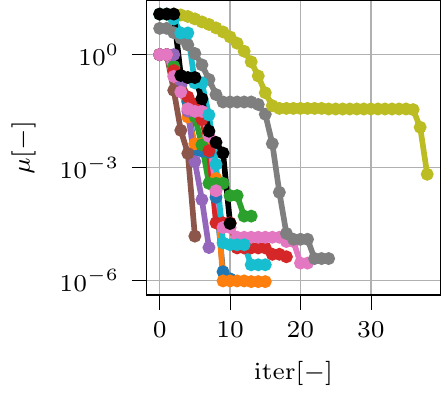}}
\caption{Development of (a) residual error $\reserr$, (b) gradient based error $\graderr$ and (c) regularization parameter $\mu$ for inverse analysis runs with different initial guesses over iterations, with Gaussian noise with standard deviation $\sigma=10^{-4}\sunit{mm}$ added to the measurements.}
\label{fig:FSIerrorse-4}
\end{figure}

\begin{figure}[htbp]
\centering
\subfloat[\label{subfig:FSIreserre-3}]{
\includetikz{width=0.3\textwidth}{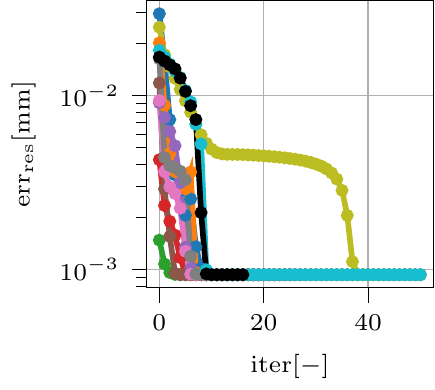}}
\hfill
\subfloat[\label{subfig:FSIgraderre-3}]{
\includetikz{width=0.303\textwidth}{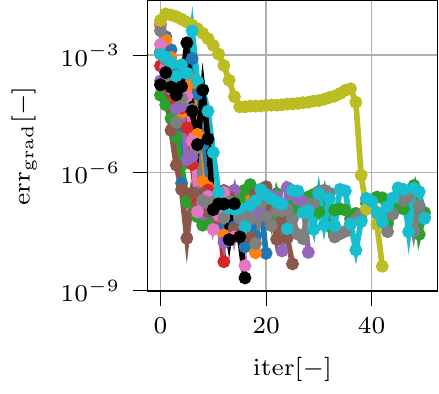}}
\hfill
\subfloat[\label{subfig:FSImue-3}]{
\includetikz{width=0.311\textwidth}{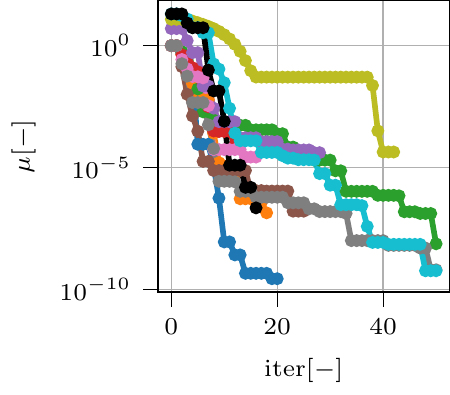}}
\caption{Development of (a) residual error $\reserr$, (b) gradient based error $\graderr$ and (c) regularization parameter $\mu$ for inverse analysis runs with different initial guesses over iterations, with Gaussian noise with standard deviation $\sigma=10^{-3}\sunit{mm}$ added to the measurements.}
\label{fig:FSIerrorse-3}
\end{figure}

\begin{figure}[htbp]
\centering
\subfloat[\label{subfig:FSIreserre-2}]{
\includetikz{width=0.3\textwidth}{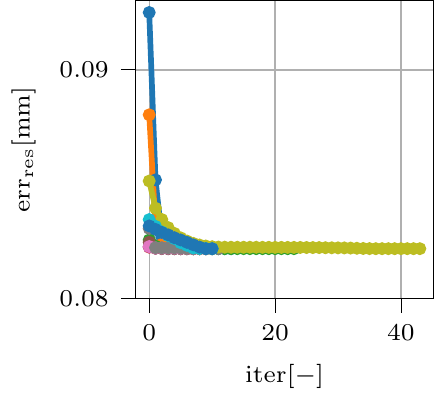}}
\hfill
\subfloat[\label{subfig:FSIgraderre-2}]{
\includetikz{width=0.31\textwidth}{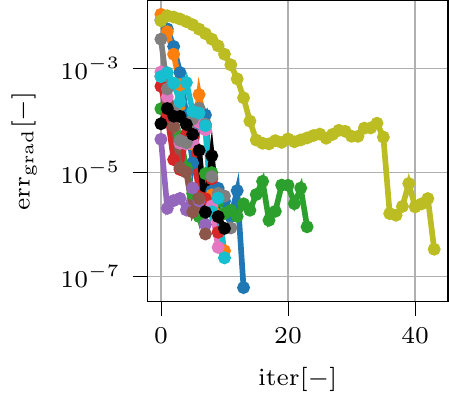}}
\hfill
\subfloat[\label{subfig:FSImue-2}]{
\includetikz{width=0.31\textwidth}{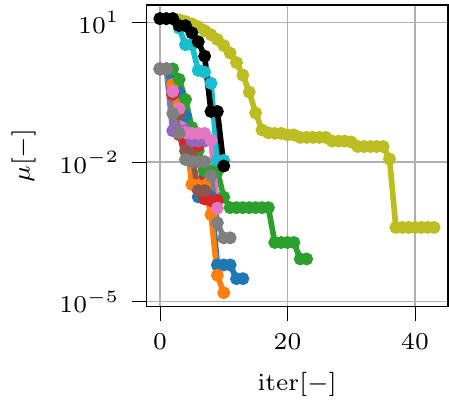}}
\caption{Development of (a) residual error $\reserr$, (b) gradient based error $\graderr$ and (c) regularization parameter $\mu$ for inverse analysis runs with different initial guesses over iterations, with Gaussian noise with standard deviation $\sigma=10^{-2}\sunit{mm}$ added to the measurements.}
\label{fig:FSIerrorse-2}
\end{figure}

From the residual and gradient based errors over the iterations, plotted in \autoref{fig:FSIerrors}, it can be seen that those quantities decrease steeply for the noise free case, when the parameters come close to the local optimum. Only the step size used in \eqref{eq:perturbation} for the finite difference approximation limits the accuracy in this setting. For the noisy data it can be seen in \autoref{fig:FSIerrorse-4} that there is a clear limit to the achievable residual errors $\reserr$ and some diffuse barrier for the gradient based errors $\graderr$ already for the slightest noise. To show this effect the convergence criteria were intentionally not adapted, although it is obvious from \autoref{fig:FSIerrorse-3} for $\sigma=10^{-3}\sunit{mm}$  that $\lmgradcrit=10^{-6}$ would have been a good choice, because there is a distinct level for $\graderr$ that cannot be reached even with many more iterations due to a low convergence criterion $\lmgradcrit$. For the data set with largest noise level shown in \autoref{fig:FSIerrorse-2} the gradient based error couldn't be reduced to a tight convergence criterion of $\lmgradcrit=10^{-8} $, but an adapted value of $\lmgradcrit=10^{-6} $ did lead to convergence for all initial guesses. It can be further observed that the numbers of iterations until a feasible level of $\graderr$ is reached does not vary much for the same initial guesses. From the third column of graphs in \autoref{fig:FSIerrors}-\autoref{fig:FSIerrorse-2}, that show the regularization parameter, it can be observed, that as expected the regularization adapts towards low values close to the result, to accelerate the progress of the iterations for all inverse analysis runs.

The path in parameter space and therefore the assumed overall shape of the residual error $ \reserr$ in parameter space does not change significantly for higher noise levels, as seen in \autoref{fig:FSIpath}, although the level of remaining residual error changes in orders of magnitude. Nevertheless, the reached optimum changes significantly for the Young's modulus $ \youngs = 506.2 \sunit{Pa} $ instead of $400\sunit{Pa} $ for the maximal noise level, but is rather insensitive for the Poisson's ratio. It is obvious and also shown in  \autoref{fig:FSIerrorse-2} that for the data set with the highest noise the level of remaining error is also the highest, but the algorithm still converges to the same local optimum for all chosen initial guesses. Mind that error plots are all presented in logarithmic scaling. So although there is far less progress in the residual error for noisy data, the algorithm finds the shifted optimum repeatedly for all initial guesses.

Looking at the "olive" green and gray lines in \autoref{fig:FSIpath} and \autoref{fig:FSIerrors} it can also be observed, that the residual error is low and rather immobile in a local vicinity of the starting points with low Young's moduli and significantly negative Poisson's ratios. Especially for the initial guess of $ \youngs = 400 \sunit{Pa} $, $ \poissonratio = -0.9$, i.e. the "olive" green line, convergence is inhibited by this indifferent shape of the residual error for low Young's modulus $ \youngs $ and negative Poisson's ratios $\poissonratio < 0$. But there is a physical explanation for this as with this type of interface location based measurement, quite similar surfaces can be obtained via large bending due to a low value for E or by lateral deformation resulting from negative $ \poissonratio$.

\subsection{Heterogeneous biofilm}

\begin{figure}[h!]
\begin{minipage}[t]{0.3\textwidth}
\centering
\includegraphics[width=\textwidth]{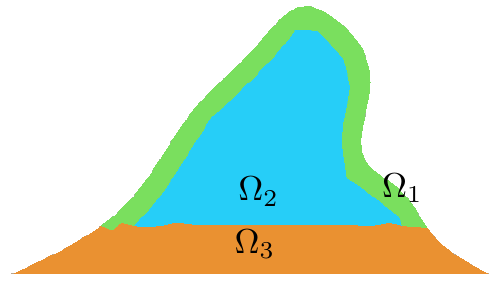}
\caption{Subdomains for heterogeneous example.}
\label{fig:subdomains}
\end{minipage}
\hfill
\begin{minipage}[t]{0.65\textwidth}
\centering
\includegraphics[width=0.46\textwidth]{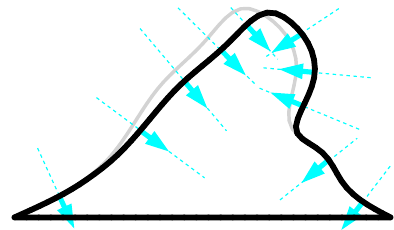}
\caption{Measurement points and directions as arrows indicating the positive directions for heterogeneous FSI example on the deformed geometry of the reference result (black) and initial geometry (gray).}
\label{fig:FSIarrowshetero}
\end{minipage}
\end{figure}

As indicated in the literature \cite{bio_stoodley2002, bio_klapper2010, bio_wang2010} biofilm material properties depend on growth regimes, age and on induced flow rates. This implicates a layer like structure that a biofilm might develop under varying conditions. An arbitrary showcase example model, with subdomains as depicted in \autoref{fig:subdomains}, is used to show that the presented method is applicable to determine different material parameters for different subdomains.

The material parameters in this reference solution are: $\youngs_1 = 500 \sunit{Pa}$, $\poissonratio_1 = 0.2$, $\youngs_2 = 200 \sunit{Pa}$, $\poissonratio_2 = 0.1 $, $\youngs_3 = 1000 \sunit{Pa}$, $\poissonratio_3 = 0.3$ in a Saint-Venant-Kirchhoff material model. The geometry used for the homogeneous biofilm model is reused here. The channel geometry and fluid volume inflow are controlled in the same way. The deformation of the biofilm under given load due to the interacting fluid forces can be seen in \autoref{fig:FSIarrowshetero}.

Measurement points and directions are also introduced in \autoref{fig:FSIarrowshetero} on the deformed geometry. It must be taken care to measure all the influences of all subdomains and therefore two measurement points were chosen on the interface of the stiffer footing layer. Measurement was conducted normal to the observed interface.

\begin{figure}[htbp]
\centering
\subfloat[\label{subfig:FSIhreserr}]{
\includetikz{width=0.3\textwidth}{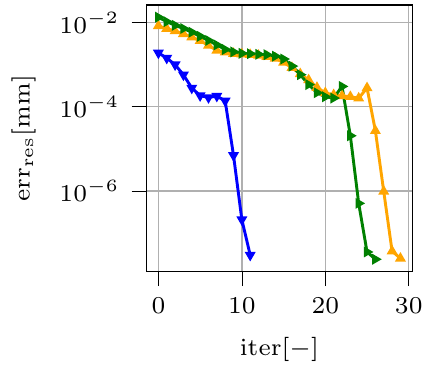}}
\hfill
\subfloat[\label{subfig:FSIhE}]{
\includetikz{width=0.304\textwidth}{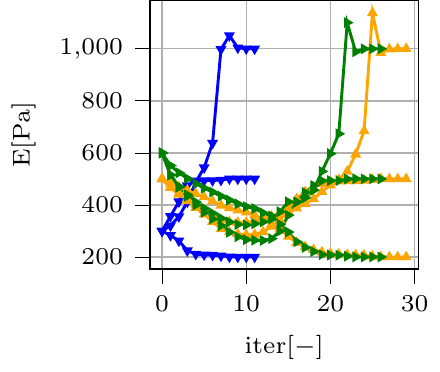}}
\hfill
\subfloat[\label{subfig:FSIhnue}]{
\includetikz{width=0.3\textwidth}{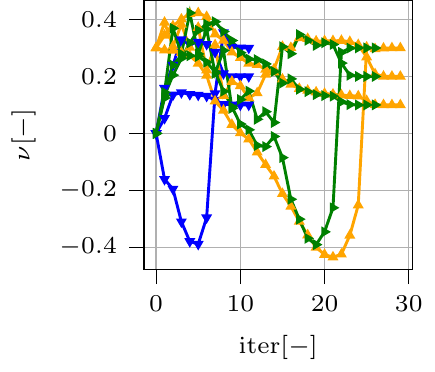}}
\caption{Result of Inverse analysis for noise free data. Development of (a) residual error $\reserr$, (b) Young's modulus $\youngs$, (c) Poisson's ratio $\poissonratio $ over iterations for different initial guesses.}
\label{fig:FSIhresult}
\end{figure}

At first it is verified that with the noise free artificial measurement data, the method allows recovering the correct material parameters. For that a short summary of algorithm runs with different initial guesses listed in \autoref{tab:initial_guesses_hetero} is plotted in \autoref{fig:FSIhresult} with the respective colors. Resulting search paths for number of parameters $\numparam > 2$ can no longer be interpreted visually with respect to the response surface in $\reserr$. Therefore search paths are plotted individually for the parameters. For these plots one color codes one inverse analysis run. In \autoref{fig:FSIhresult} it can be observed that the inverse analysis is able to find the reference parameters for noise free data for different initial guesses. It was observed that this problem converges faster if Young's modulus $\youngs$ is underestimated in the initial guess. These algorithm runs are the ones with the highest number of parameters presented. It is obvious in the plots, that the search paths for all parameters are interdependent. This is also expected from the algorithm as for every iteration a finite difference approximation of the partial derivatives in all parameters is used to find the next step. The great increase in convergence speed seen in \autoref{subfig:FSIhreserr} below $\reserr = 10^{-4}\sunit{mm}$ is a hint, that a very local optimum is found, as also the parameters do not change much for those respective last iterations.

\begin{table}[htbp]
\caption{Initial guesses for different inverse analysis runs in heterogeneous example.}
\centering
\input{table_initial_guesses_hetero.tex}
\label{tab:initial_guesses_hetero}
\end{table}

Real OCT resolution is in the range of $\unit{\mu m} $ \cite{kit_bio_wagner2017, kit_bio_gierl2020}, so the following examples will be run after Gaussian noise with standard deviation $10^{-3}\sunit{mm}$ was added to the generated artificial measurement data.

\begin{remark}[Estimation of initial guess]
\label{rem:heterohomo}
For this setting several arbitrary initial guesses did not lead to convergence of the method for the noisy data. That is why the problem is first run with a reduced set of parameters. To do so and set an application oriented scenario, where the heterogeneous character is unknown and the individual parameters are unknown, the domain is assumed to be homogeneous and the material parameters that fit that assumption are searched for. This also helps to loosen the strong one-to-one relationship between generated data and forward model evaluations. The results are displayed in \autoref{fig:FSIhhresult}.

\begin{figure}[htbp]
\centering

\end{figure}

\begin{figure}[h!]
\begin{minipage}[t]{0.65\textwidth}
\centering
\subfloat[\label{subfig:FSIhhE}]{
\includetikz{width=0.44\textwidth}{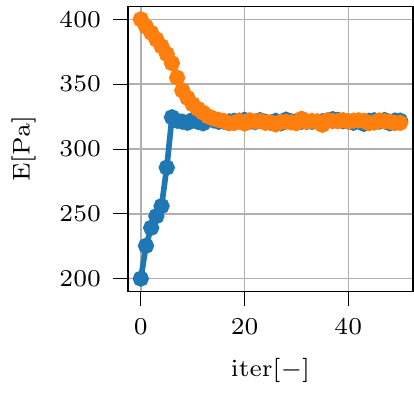}}
\hfill
\subfloat[\label{subfig:FSIhhnue}]{
\includetikz{width=0.46\textwidth}{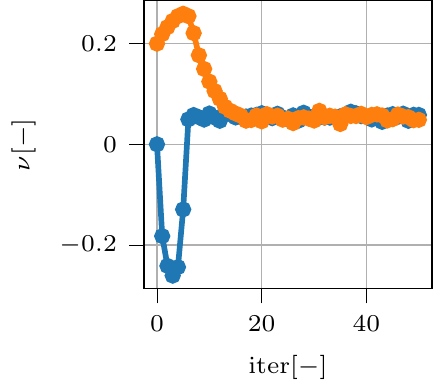}}
\hspace{2mm}
\caption{Result of inverse analysis for added Gaussian noise with $\sigma = 10^{-3} \sunit{mm} $. (a) Young's modulus $\youngs$ (b) Poisson's ratio $\poissonratio $ for two different initial guesses.}
\label{fig:FSIhhresult}
\end{minipage}
\hfill
\begin{minipage}[t]{0.35\textwidth}
\captionsetup{type=table}
\caption{Initial guesses for different inverse analysis runs in heterogeneous example with simplification to homogeneous assumption.}
\centering
\input{table_initial_guesses_hetero_homo.tex}
\label{tab:initial_guesses_hetero_homo}
\end{minipage}
\end{figure}
\begin{figure}[htbp]
\centering
\subfloat[ \label{subfig:FSIhhreserr}]{
\includetikz{width=0.3\textwidth}{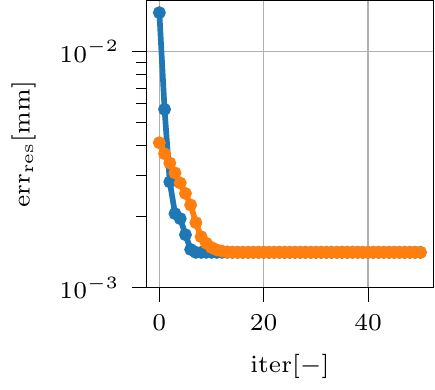}}
\hspace{1cm}
\subfloat[\label{subfig:FSIhhgraderr}]{
\includetikz{width=0.3\textwidth}{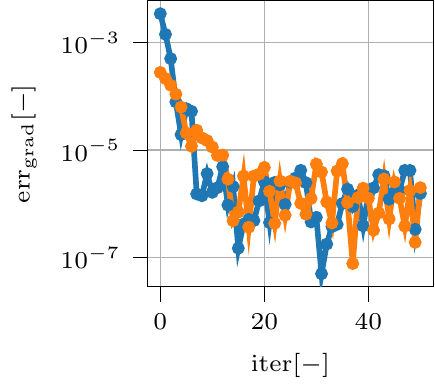}}
\caption{Errors for inverse analysis with homogeneous assumption on heterogeneous reference results with added Gaussian noise with $\sigma = 10^{-3} \sunit{mm} $. (a) residual error $\reserr$, (b) gradient based error $\graderr$ for two different initial guesses. }
\label{fig:FSIhherrors}
\end{figure}

In \autoref{fig:FSIhherrors} it shows that the convergence criterion of $ \graderr < 10^{-9} \unit{mm}$ could not be met. So the result must somehow be concluded from the iterations. It is recommended to judge by an adjusted, but still objective convergence criterion. Therefore the data criterion is adjusted to $ \graderr < 10^{-6} \unit{mm}$. With adjusted convergence criterion the averaged result for two different initial pairs of values, listed in \autoref{tab:initial_guesses_hetero_homo}, is $\youngs = 323 \sunit{Pa} $ and $ \poissonratio = 0.0656 $. For $ \youngs $ that is well in between the parameters used for the reference data and for $ \poissonratio $ that is below all the original values. From this simple numerical experiment we can already conclude that the achievable level of gradient based error $\graderr $ cannot be predicted and hence the convergence criteria must be tuned to the data used. Nevertheless the result achieved with the algorithm is conclusive even if the algorithm did not converge under too strict criteria.
\end{remark}

In the next step it can be assumed that the domain is in fact heterogeneous and $ \poissonratio $ is equal for all subdomains. To set this example $\poissonratio_1=\poissonratio_1=\poissonratio_1=0.1 $ is rounded from the result with the assumption of a homogeneous domain in \autoref{rem:heterohomo} and $\youngs=300 \sunit{Pa} $ is picked as an initial guess for an inverse analysis for $\youngs$ in the three subdomains. This results in a distribution of $\youngs_1 = 497 \sunit{Pa}$, $ \youngs_2 = 226 \sunit{Pa}$, $ \youngs_3 = 510 \sunit{Pa} $. The correct tendency in stiffness in the subdomains is apparent. Only in the footing layer the result is far off the reference value. Most likely the influence of the stiffness of the footing layer is not conclusive enough towards the shape based surface comparison. The remaining residual error is $\reserr = 8.98\cdot10^{-4} \sunit{mm} $. That is lower than the solution with the homogeneous approach and even lower than the residual for a model evaluation with the parameters used for the noise free reference result. This means the added noise has altered the data in a way, that it does no longer represent the reference result in the position of the optimum. If as a further step this result is used as an initial guess for a new optimization for all six Parameters, the residual error can be lowered to $\reserr = 7.43\cdot10^{-4} \sunit{mm} $ with the result $\youngs_1 = 273 \sunit{Pa}, \poissonratio_1 = -0.60$, $\youngs_2 = 192 \sunit{Pa}, \poissonratio_2 = -0.10 $, $\youngs_3 = 491 \sunit{Pa}, \poissonratio_3 = 0.39$. It appears this type of problem and the measurement only via interface deformation favors the assumption, that the material is auxetic, i.e. $\poissonratio < 0.0 $. It appears that in this numeric experiment the field of residual error has flipped and the optimum has shifted towards softer, auxetic materials. It is a further conclusion that optimizing for $\youngs$ only is more robust, than the combination of both $ \youngs$ and $ \poissonratio $ at once because a negative $ \poissonratio $ can compensate a $ \youngs $ that is too low in the surface measure. That means that lateral expansion will fill the gap to the optimum for too much bending. Auxetic materials are very rare and mostly occur in specially designed materials with very unique microstructures. As long as that is not proven for the material of interest it is a valid assumption that $ \poissonratio $ is positive.

It is known that inverse analysis not only depends on the physical problem at hand as well as on the forward models used, but also on the type and amount of measurement information. For the above type of problems, simply more and other measurement input would be needed in order to allow for identification of all values for Young's modulus and Poisson ratio at the same time.

\subsection{Two-phase poroelastic biofilm}

\begin{figure}[htbp]
\centering
\subfloat[\label{subfig:FPIdisp}]{
\includegraphics[width=0.49\textwidth]{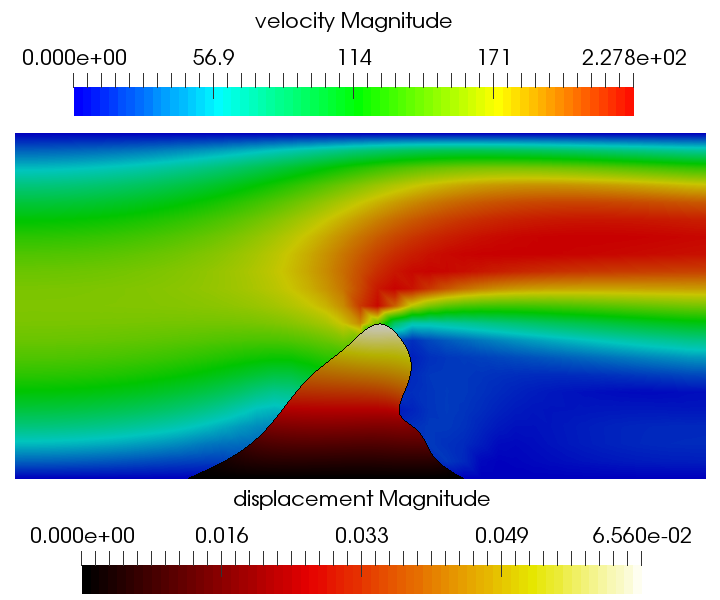}}
\hfill
\subfloat[\label{subfig:FPIporos}]{
\includegraphics[width=0.49\textwidth]{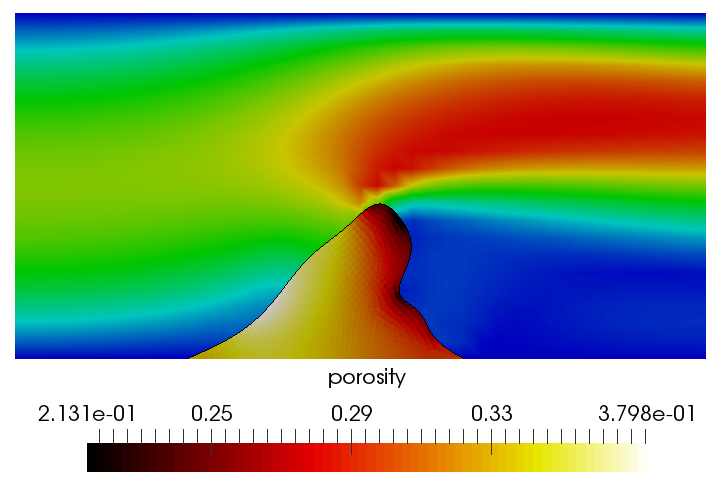}}
\caption{Solution of the reference simulation for the porous example. Fluid velocity magnitude and (a) skeleton displacement and (b) porosity.}
\label{fig:FPIdisp}
\end{figure}

\begin{figure}[htbp]
\centering
\subfloat[\label{subfig:FPIvel}]{
\includegraphics[width=0.49\textwidth]{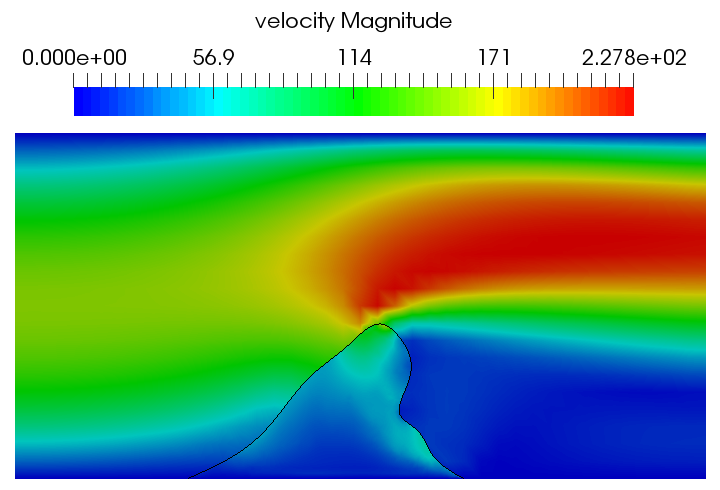}}
\hfill
\subfloat[\label{subfig:FPIpres}]{
\includegraphics[width=0.49\textwidth]{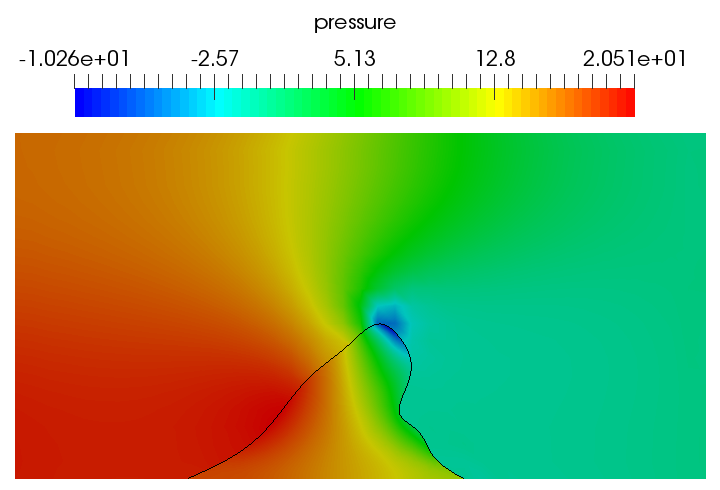}}
\caption{Results for fluid velocity (a) and fluid pressure (b) of reference simulation illustrated on the deformed geometry both for "external" flow field and fluid inside the porous medium.}
\label{fig:FPIresult}
\end{figure}

\begin{figure}[htbp]
\centering
\includegraphics[width=0.3\textwidth]{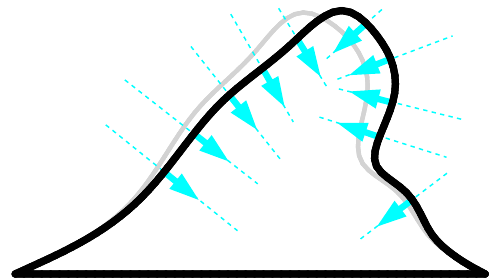}
\caption{Measurement points and directions as arrows indicating the positive directions for poroelastic biofilm example on the deformed geometry of the reference result (black) and initial geometry (gray).}
\label{fig:FPIarrows}
\end{figure}

The porous nature of biofilms is well documented. Measurements from OCT scans allow an estimation of biofilm porosity \cite{kit_bio_wagner2017}. In the following the attempt to determine porosity via inverse analysis will be presented. The reference result that serves as dummy experiment is set up in a similar manner to previous FSI examples. The biofilm and channel geometry are the same. We switch to the quasi two-dimensional setting with thickness $ 0.01 \sunit{mm}$ and use the same inflow rate $100\sunitfrac{mm^2}{s}\cdot0.01\sunit{mm} = 1\sunitfrac{mm^3}{s} $ over the height of the left boundary. All displacements and velocities in thickness direction are restricted to zero. Further parameters to the fluid-poroelasticity interaction are arbitrarily chosen as permeability $\matpermeabpscalar = 10^{-4} \sunit{{mm}^2} $ and Poisson's ratio $\poissonratio = 0.3 $. The fluid phase in the poroelastic domain also has the properties of water. For the reference result the parameters $ \youngs = 300 \sunit{Pa}$ and a homogeneous initial porosity of $\initial{\porosity} = 0.25$ are used in a Neo-Hookean material model. As we are using a fully coupled two-phase poroelastic model, porosity changes due to deformations caused by interaction forces from the external flow field but also due to pressure within the porous medium itself. The solution for the velocity magnitude, displacement magnitude and porosity of the biofilm are shown in \autoref{fig:FPIdisp}. It is observed that the biofilm bends with the flow to the right. This is depicted in \autoref{fig:FPIarrows} where the initial and deformed geometry ale plotted on top of each other. The porosity opens up in the stretched upstream part and is reduced in compressed downstream regions (see \autoref{subfig:FPIporos}). The results for velocity and pressure solution of the fluid, inside and outside of the biofilm, are depicted in \autoref{fig:FPIresult}.

\begin{figure}[h!]
\begin{minipage}[c]{0.5\textwidth}
\centering
\includetikz{width=0.7\textwidth}{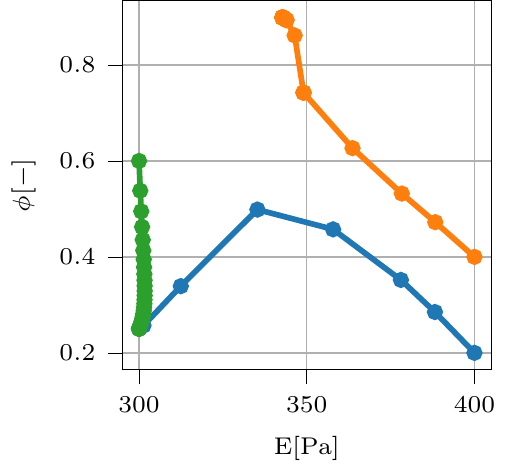}
\caption{Path in parameter space towards optimum for porous example.}
\label{fig:FPIpath}
\end{minipage}
\hfill
\begin{minipage}[c]{0.45\textwidth}
\captionsetup{type=table}
\caption{Initial guesses for different inverse analysis runs in two-phase poroelastic example.}
\centering
\input{table_initial_guesses_poro.tex}
\label{tab:initial_guesses_poro}
\end{minipage}
\end{figure}

The measurement points used for the inverse analysis are displayed in \autoref{fig:FPIarrows} on the deformed geometry. They are chosen where the most significant deformation of the interface shape is expected. Knowing, that the varying porosity is coupled to the biofilm deformation one measurement point is chosen on the lower right bump of the geometry. On that basis the inverse analysis is conducted for different initial guesses listed in \autoref{tab:initial_guesses_poro}. The paths in parameter space are shown in \autoref{fig:FPIpath}, wherein each marker represents one Levenberg-Marquardt iteration. For two initial guesses, namely the blue and green line, the algorithm converges to the reference parameters and for one initial guess, displayed in orange, the algorithm had to be terminated without result.

\begin{figure}[htbp]
\centering
\subfloat[\label{subfig:FPIeres}]{
\includetikz{width=0.3\textwidth}{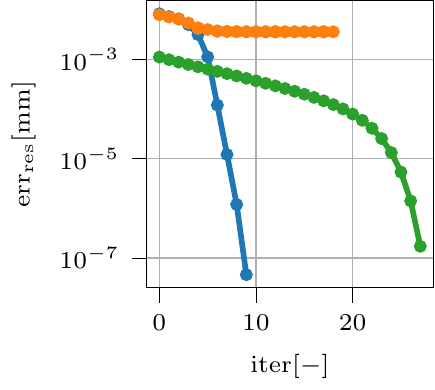}}
\hspace{1cm}
\subfloat[\label{subfig:FPIegrad}]{
\includetikz{width=0.3\textwidth}{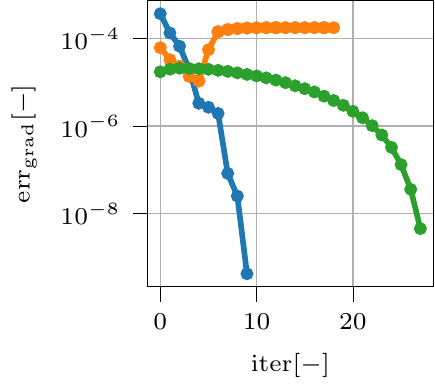}}
\caption{Result of inverse analysis for porous setting (a) residual error $\reserr$ (b) gradient based error $\graderr$ over iterations. }
\label{fig:PFIresult}
\end{figure}

It appears that the measured deformation of the interface is not fully conclusive towards the biofilm material porosity, as higher porosity, due to the interplay between porosity and Young's modulus, also lowers the effective stiffness of the porous medium. And as the porosity is naturally bounded $ \porosity \in [0,1] $, the algorithm tends towards the upper bound. Since the gradient indicates further improvement towards this bound, the algorithm gets stuck in the applied upper bound of $ \imax{\porosity} = 0.9 $  and the upper limit for the adaptive regularization parameter terminates iterations. Nevertheless, if the initial guess for Young's modulus is good enough, the optimum can still be found, although it takes many steps. It can be observed in \autoref{fig:PFIresult}, that also the convergence speed depends strongly on the path in parameter space and therefore obviously also on the quality of the initial guess. Along the first (blue) path the analysis converges in nine steps, whereas the one with a higher starting value for the porosity (green) takes 27.

It would also be desirable to include more parameters into the inverse analysis and, for example, to additionally optimize for permeability $ \matpermeabpscalar $ or Poisson's ratio $\poissonratio$ in the same algorithm. However this short example is only meant to serve as a proof of concept for the suggested approach. In case many parameters and their interplay need to be considered, it definitely makes sense to also include sensitivity analysis and probably also consider probabilistic based (inverse) analysis approaches.

\subsection{Surface growth of biofilm}
\label{subsec:growth}

This last example demonstrates that the method is also applicable to identify parameters in growth models as the one developed and used in \cite{lnm_bio_coroneo2014}. The inflow rate $ 0.1 \sunitfrac{mm^2}{s} \cdot0.01\sunit{mm} = 10^{-3}\sunitfrac{mm^3}{s}$, that is induced from the left side, is much lower in this example as it is applied over a long time period providing growth conditions for the biofilm. Growth processes take place on a different time scale than dynamic FSI and this is accounted for in the temporal multi-scale approach detailed in \cite{lnm_bio_coroneo2014}. The geometry of the problem is chosen as the one presented in \cite{lnm_bio_coroneo2014} to sustain comparability. The flow channel has the dimensions $ 0.6\sunit{mm} \times 0.3 \sunit{mm} \times 0.01\sunit{mm}$ in a quasi twodimensional model with no displacement or velocity in thickness direction. The biofilm is represented by a finger like structure of $0.04 \sunit{mm} $ width and $0.1\sunit{mm}$ height that ends up with a semi circular tip. The inflow velocity profile is parabolic according to the other examples. The FSI time period is $5\sunit{s}$ and the fluid inflow rate is increased smoothly. The growth time period is one day. The material parameters for the Saint-Venant-Kirchhoff material are $\youngs = 100\sunit{Pa} $ and $ \poissonratio = 0.3$. They are assumed to be known from a deformation experiment. Like in \cite{lnm_bio_coroneo2014} the concentration of the scalar species at the inflow is chosen in the range of oxygen dissolved in water as $ \conc^\fluidletter_\mathrm{in} = 2.5\cdot 10^{-11} \sunitfrac{mol}{mm^3}$. The reaction rates in \eqref{eq:monod} are assumed as $ K_1^\mathrm{R} = 3.0\cdot 10^{-11} \sunitfrac{mol}{(mm^3\,s)} $, $K_2^\mathrm{R} =  3.0\cdot 10^{-12} \sunitfrac{mol}{mm^3}$. The Diffusion coefficient for both phases is $\diffusivity^\structletter = \diffusivity^\fluidletter = 2.5\cdot 10^{-3} \sunitfrac{mm^2}{s}$. The scalar in the scalar transport problem represents a dummy nutrient for the biofilm and will be referred to as such in the following.

\begin{figure}[htbp]
\centering
\subfloat[\label{subfig:growthvelbend}]{
\includegraphics[width=0.49\textwidth]{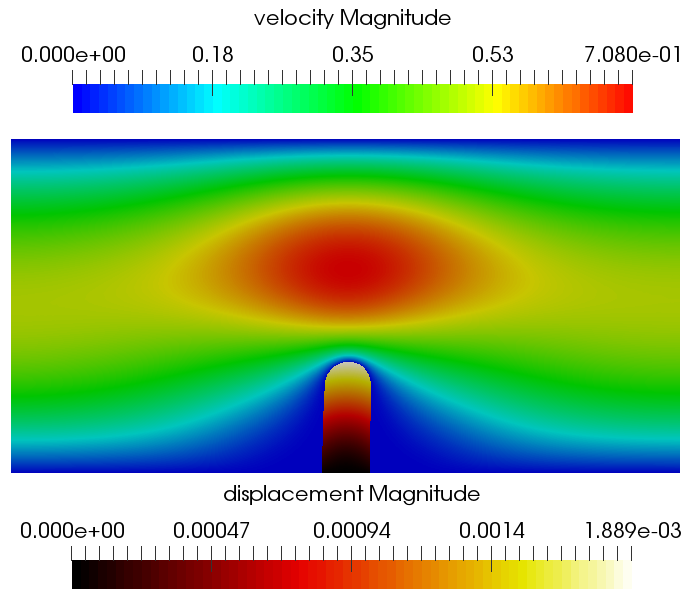}}
\hfill
\subfloat[\label{subfig:growthpressurfgr}]{
\includegraphics[width=0.49\textwidth]{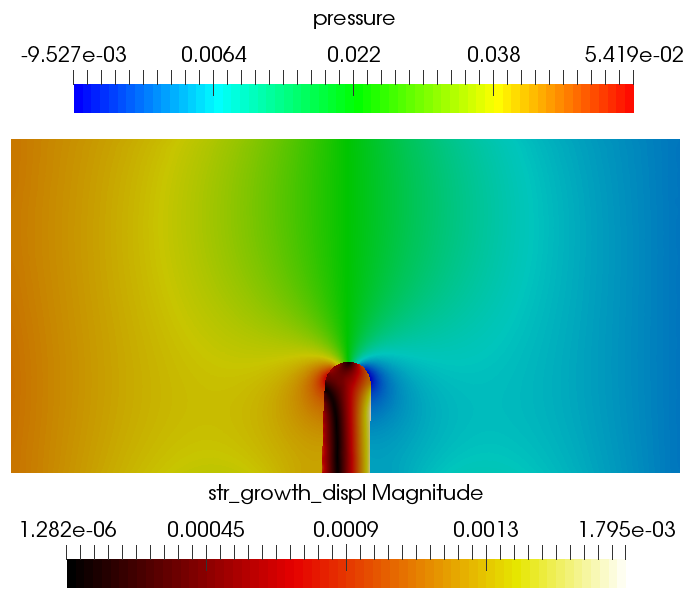}}
\caption{Solution of the reference simulation illustrating the flow field and the deformed biofilm structure. (a) Fluid velocity and solid displacement field (b) Fluid pressure and growth magnitude on interface as well as ALE displacements within biofilm domain.}
\label{fig:growthresult}
\end{figure}

\begin{figure}[h!]
\begin{minipage}[b]{0.49\textwidth}
\centering
\includegraphics[width=\textwidth]{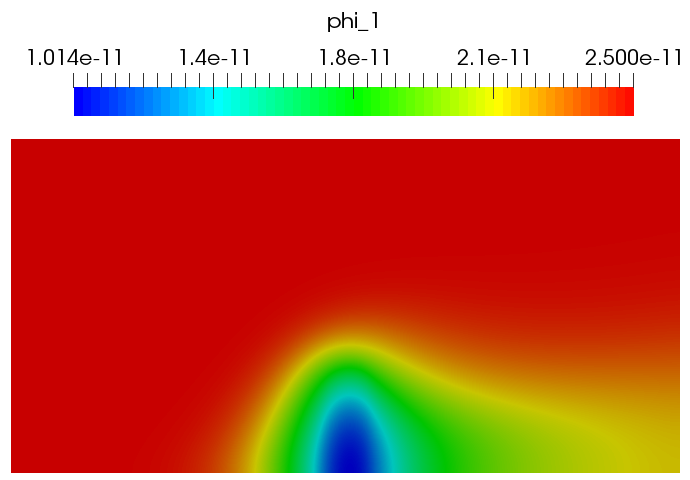}
\caption{Solution of the prototype nutrient concentration distribution.}
\label{fig:growthscatraresult}
\end{minipage}
\hfill
\begin{minipage}[b]{0.5\textwidth}
\centering
\includegraphics[width=0.32\textwidth]{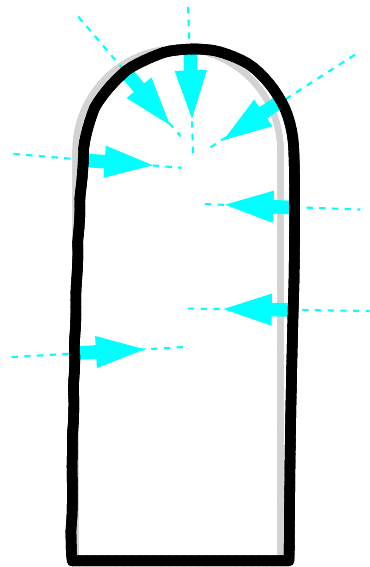}
\caption{Measurement points and directions as arrows indicating the positive directions for biofilm example with surface growth on the actual grown and deformed geometry of the reference result (black) and initial geometry (gray).}
\label{fig:Garrows}
\end{minipage}
\end{figure}

Model equation \eqref{eq:3compgrowthlaw} shows that the used growth model depends on three different parameters, namely $K_1^\growthindex $ as the factor for growth on the domain boundaries scaling with the actual nutrient flux, $K_2^\growthindex $ as the factor for inhibition of growth due to normal stresses and $K_3^\growthindex $ as the factor for inhibition of growth due to shear stresses. They all appear linearly in the chosen surface growth model. For the reference result the parameters $K_1^\growthindex = 6\cdot10^{4}\sunitfrac{mm^3}{mol}$, $K_2^\growthindex=5\cdot10^{-2}\sunitfrac{mm^2\,s}{g} $ and  $K_3^\growthindex=8\cdot10^{-2}\sunitfrac{mm^2\,s}{g} $ were used. The reference solution is shown in \autoref{fig:growthresult} for the fluid velocity and pressure and displacements due to hyperelastic deformation and growth and for the scalar transport species concentration in \autoref{fig:growthscatraresult}. The deformed (bended and grown) biofilm and the initial shape are plotted in \autoref{fig:Garrows}.

The changes in biofilm geometry due to displacements and due to growth range in the same order of magnitude. In \autoref{subfig:growthpressurfgr} surface growth is displayed at the interface along with the solid ALE field on the biofilm domain. This gives a more intuitive overview of the growth deformation, although the ALE field has no physical meaning and the plotted growth stems from a pure surface growth model. The nutrient that is transported through the flow is consumed in the biofilm domain according to the reaction rate in \eqref{eq:monod}. This produces a gradient over the interface and therefore nutrient flux, that leads to growth. On the upstream side the growth because of nutrient supply and the inhibition of growth because of higher tractions are more balanced, whereas on the downstream side, where the fluid induces lower tractions, a larger growth is observed at the interface, although the nutrient flux is lower. Over the finger tip the erosive effects of the traction can be observed.

The interface deformations are measured in the points displayed in \autoref{fig:Garrows} on the fully deformed geometry. The distribution of the measurement points is based on the anticipated different regimes - for growth and FSI - on the upstream, downstream and tip side. There must be points in regions with large growth resulting from high nutrient flux or low interface tractions, in regions with high tangential components of the interface tractions and regions with high normal components of the interface tractions.

\begin{figure}[htbp]
\centering
\subfloat[\label{subfig:growthk1}]{
\includetikz{width=0.305\textwidth}{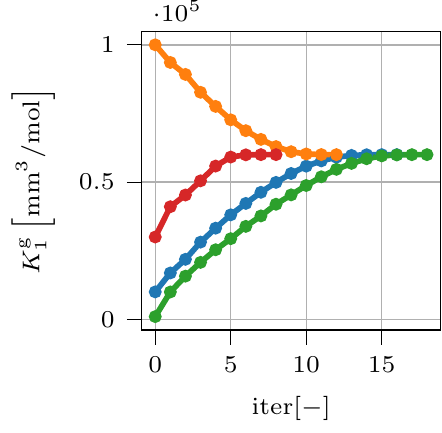}}
\hfill
\subfloat[\label{subfig:growthk2}]{
\includetikz{width=0.3\textwidth}{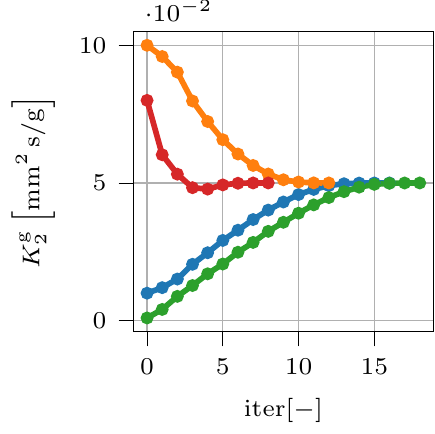}}
\hfill
\subfloat[\label{subfig:growthk3}]{
\includetikz{width=0.3\textwidth}{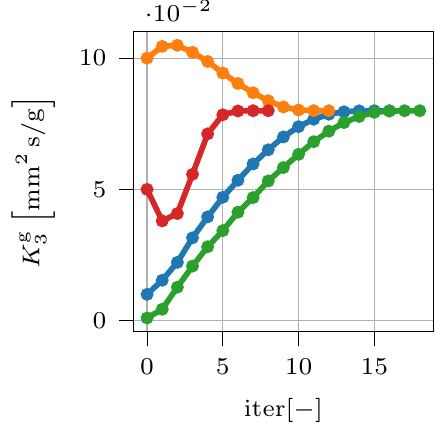}}
\caption{Result of inverse analysis over iterations for growth parameters (a) $K_1^\growthindex $, (b) $K_2^\growthindex $, (c) $K_3^\growthindex $.}
\label{fig:growthparams}
\end{figure}

\begin{figure}[htbp]
\centering
\subfloat[\label{subfig:growth_eres}]{
\includetikz{width=0.3\textwidth}{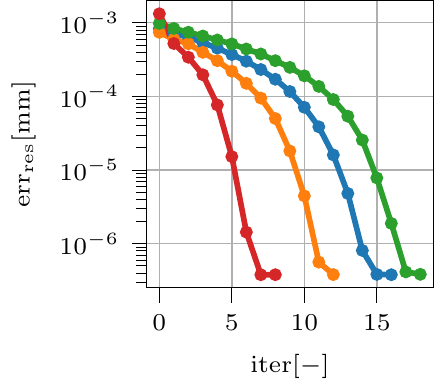}}
\hspace{1.0cm}
\subfloat[\label{subfig:growth_egrad}]{
\includetikz{width=0.304\textwidth}{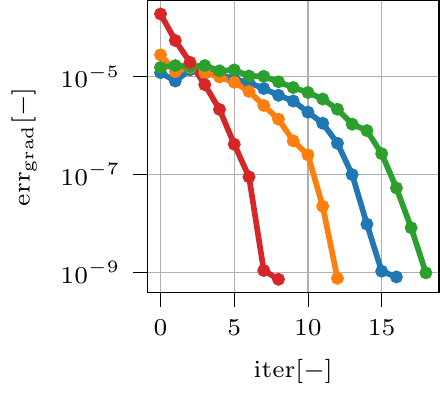}}
\caption{Errors over iterations for growth inverse analysis (a) residual error $\reserr $ (b) gradient based error $\graderr $.  }
\label{fig:growtherrors}
\end{figure}

Initial guesses used for the inverse analysis of the growth parameters are listed in \autoref{tab:initial_guesses_growth} and results obtained are plotted in \autoref{fig:growthparams}. The errors plotted in \autoref{fig:growtherrors} decline over the optimization iterations. It is observed that the search path depends on all growth parameters at once. That means that the residual error $\reserr$, as the norm over surface distances, does not measure the three contributions of the parameters to growth and erosion independently in this analysis. This problem setting allows to have initial guesses further away from the optimum, as long as growth is significantly smaller than the dimensions of the biofilm domain. The parameters from the reference data are the result of the inverse analysis for all analyzed initial guesses.

\begin{table}
\centering
\captionsetup{type=table}
\caption{Initial guesses for different inverse analysis runs in growth example.}
\centering
\input{table_initial_guesses_growth.tex}
\label{tab:initial_guesses_growth}
\end{table}

\section{Discussion}
\label{sec:discussion}

Beside the discussion on presented examples also some general remarks and discussions regarding general aspects of the approach that cannot be shown in examples are added for the completeness of the presentation. Further, anticipated general aspects regarding the application of the method are summarized.

\subsection{General discussion of the approach}

\paragraph{Significance of parameters} We have looked at coupled models, that displayed different shapes of residual errors over model iterations in parameter space with the given measure of the biofilm surface shape. In several combinations, compensation effects in the parameters occurred (e.g. $\youngs $ and $\poissonratio $ or $ \youngs$ and $\porosity$). The presented approach can only be used if the information to all parameters analyzed are actually at least somehow represented in the data and also show effect in the biofilm surface shape in at least partially independent patterns. For presented physical biofilm models the key parameters could nevertheless be determined in example inverse analyses.

\paragraph{Model response} Presented method is not designed to explore the full parameter space, which would be interesting for the global view on the plausibility of combinations of parameters in the whole parameters space for given data. For exploring the residual error on an interval in the parameter space there is a great variety of random or quasi random sampling methods. To efficiently get an global estimation of the objective function in the full parameter space regression methods like for example Gaussian processes \cite{stat_rasmussen2005} on respective sample results can be applied. The advantage of the presented method is, that it is not necessary to explore the whole parameter space but to find a local minimum with a limited number of iterations and therefore limited amount of forward model evaluations.

\paragraph{Deterministic character} It should be emphasized that the Levenberg-Marquardt optimization is a deterministic approach and there is a risk that the numerical model used for an analysis is not stable along the full search path and especially in the vicinity of the optimum. If the forward model cannot be evaluated and therefore no measurable results retrieved, the finite differences cannot be computed and the algorithm must be terminated, as there is no strategy for following iterations. This drawback further restricts the choice of initial guesses to a set of parameters with which the forward model can be solved.

\paragraph{Computational cost} The computational cost of the algorithm is dominated by the cost for the forward model evaluations. Therefore it scales at least linearly with increasing number of parameters $\numparam $, as more simulations are required per iteration for the finite difference approximation of the Jacobian. Additionally, a higher dimension in parameters of the inverse problem leads to more complex objective functions and therefore potentially longer search paths. In the same sense a higher number of parameters $ \numparam$ will potentially lead to a smaller region around the optimum for the individual parameters from which the initial guess has to be chosen to be able to find an optimum. In presented examples the algorithm converged or failed in 20-40 iterations.

\paragraph{High number of parameters} From all of the above points it becomes clear that the presented method does not scale arbitrarily well for high number of parameters. The more parameters involved in the model, the more likely it is, that they are differently significant to the forward model solutions and possibly interact in their contribution to the objective function. The objective function gets more complex with higher number of parameters and the probability, that it becomes multimodal, i.e. more than one local optimum exists, increases. The more complex the objective function is, the more likely it is, that the search path leads to parameter regions, where the forward model cannot be evaluated and the inverse analysis yields no result. Higher number of parameters also leads to higher computational cost of a finite difference approximation and search paths grow longer in more complex objective functions increasing the number of finite difference approximations necessary. Overall it is expected that the cost and general applicability of the proposed method scale poorly for high parameter dimensions.

\subsection{Comments on application}

\paragraph{Importance of initial guess} The presented algorithm includes a local optimizer and it cannot find a reasonable solution for every given combination of parameters as initial guesses. The availability of a good initial guess is crucial, as it decides if the method converges and also how many mostly costly forward model evaluations are required. Nevertheless the presented algorithm itself can help to find a good initial guess, when it is used with a reduced set of parameters. In the application with real experiments it is unknown if an optimum that was found with one initial guess using the presented approach is a global optimum with regard to given data. Hence inverse analysis results must always be carefully interpreted with respect to plausibility of the results. In doubt it is always possible to validate results by using a second and significantly different initial guess and compare the results.

\paragraph{Model selection} Deterministic inverse analysis is used to find a point estimate only, representing a local best fit of parameters in a chosen forward model. It does not provide a general answer to what type of measurement error is inherent in the data and also not if the forward model used for optimization is itself a good choice or should be improved. This statement is not restricted to the material model but also holds for the physical model and the boundary conditions used for the forward model. In the case of raw flow cell experiment results for example, the best choice for a specific hyperelastic material model is unknown and a Saint-Venant-Kirchhoff material model with its two parameters is just one very simple choice. The selection of a suitable forward model is up to the analyst with presented approach, but also could be included in the inverse analysis by identifying the best parameters for different models and comparing the overall approximative error (like also done in e.g. \cite{lnm_inv_rausch2011}). If one does that it is very important to relate the approximative quality to the number of parameters in the model (e.g. by the Bayesian or the Akaike information criterion) as we are seeking a predictive model and not just a fit to some data points (see e.g. \cite{lnm_inv_rausch2011} or \cite{lnm_inv_nagler2017}).

\paragraph{OCT imaging} The presented measure used for the objective function works independent of the physical model and the spatial dimension of the model. An important feature is that it can easily be used tailored to the data gained from OCT. For example, in the unloaded state a three dimensional scan of a biofilm can be used to build a mesh for computational evaluation and the objective function can be defined with so called B-Scans from the loaded and therefore deformed biofilm, whenever speed of image acquisition is the limit to measurement quality. B-Scans are two dimensional plane scans of the flow cell \cite{kit_bio_blauert2015} and naturally faster to acquire than three dimensional images which are just stacks of multiple B-Scans.

\paragraph{Combination of data} Only examples that include information of the initial non deformed geometry of the biofilm incorporated in the forward model mesh and one quasi steady state solution are shown for the sake of concise presentation. But obviously also applications with model evaluations in different time steps or load steps to asses nonlinear material parameters in more complex material models or viscous effects of the biofilm can be easily treated. In such applications, questions of scales and weights of the contribution to the residual $ \lmres $ in the objective function must be answered, to not blindly weigh highest displacements the highest. Nevertheless it has been shown in \cite{lnm_inv_birzle_2019a} that also results from different experiments on the same specimen can be scaled and used in a single Levenberg-Marquardt based inverse analysis.

\paragraph{Time scales} In the application of the method, the experiments and results should be grouped in meaningful sets by the time scales involved. One use case can be to first analyze a deformation experiment in the flow cell and determine material parameters in a simple material model and as a second step use other experiments with a more extensive model like additional growth and determine the growth parameters (like in the example in \autoref{subsec:growth}). In this scenario the growth and hyperelastic material parameters can be regarded as decoupled as time scale for growth is orders of magnitude larger than the time scale for fluid-solid interaction under constant fluid inflow rate.

\paragraph{Convergence criterion} A convergence criterion based on combined maximal number of algorithm iterations and a maximum gradient based error presented itself as a good choice. Even with the presented artificially generated data it was not easily predictable how low the remaining residual error between measured deviation of forward model outcome and experiment observation is. Especially if artificial noise was added to the data, it showed that there was an individual distinct level of residual error, when no better set of parameters could be found. This is also obvious as the noisy data does not define a real physical solution and hence an error has to show up. This combination of convergence criteria can therefore be recommended and needs to be tuned to the specific application depending on the unknown structure of uncertainties in the experiment result data and the cost of forward model evaluations.

\section{Conclusion}
\label{sec:conclusion}

An inverse analysis method for biofilms has been presented and successfully tested on several selections of significant parameters in different aspects and models of biofilm physics. The algorithm works with a local best fit for parameter estimates in given forward models related to experimental results in the sense of least squares. A simple hands-on measure for the comparison of shapes of biofilms in deformation experiments has been presented and tested. It has been shown for a veriety of different meaningful models for biofilms, like homogeneous, heterogeneous, poroelastic and biofilm models including surface growth, that the presented approach allows the inverse analysis for multiple key parameters at once. As the presented measure for the difference of a biofilm surface between experiment and model evaluations is based purely on their shapes, the approach can without restrictions be used for further different physical models, like ones for detachment, self contact and viscoelasticity.

\section{Acknowledgements}

Funding by the German Research Foundation (DFG) with project number WA 1521/22 for this work is gratefully acknowledged. The basis version of the software QUEENS was provided by the courtesy of AdCo Engineering\textsuperscript{GW} GmbH, which is gratefully acknowledged. We thank our project partners at Karlsruhe Institute of Technology (KIT) L. Gierl, M. Wagner and H. Horn for the collaboration that has helped to develop a method that is in the end applicable to real data acquired from experiments.

\printbibliography


\end{document}

%% file: symbols.tex
\newcommand{\beaversjoseph}{{\mathrm{BJ}}}

\newcommand{\tns}[1]{\boldsymbol{#1}} 
\newcommand{\partiald}[2]{\dfrac{ \partial #1}{\partial #2}}
\newcommand{\norm}[2]{\ensuremath{\left\|#1\ensuremath\right\|_{#2}}}

\newcommand{\grad}{\boldsymbol{\nabla}}
\renewcommand{\div}{ \grad \! \cdot \!}
\newcommand{\divref}{ \initial{\grad} \! \cdot \!}

\newcommand{\abs}[1]{\ensuremath{\left|#1\right|}}

\newcommand{\identity}{\tns{I}}
\newcommand{\trace}[1]{\mathrm{tr}\left( #1 \right)}

\global\long\def\transpose#1{#1^{\mathsf{T}}}
\global\long\def\inverse#1{#1^{\mathsf{-1}}}
\global\long\def\invtrans#1{#1^{\mathsf{-T}}}

\newcommand{\evalat}[2]{\left. #1 \right|_{#2}}

\newcommand{\initial}[1]{{#1}_0}
\newcommand{\initialtop}[1]{{#1}^0}
\newcommand{\imin}[1]{{#1}_{\mathrm{min}}}
\newcommand{\imax}[1]{{#1}_{\mathrm{max}}}

\newcommand{\paramletter}{x}
\newcommand{\numparam}{{\mathrm{n}_\mathrm{\paramletter}}}
\newcommand{\numres}{{\mathrm{n}_\mathrm{r}}}
\newcommand{\paramvec}{\tns{\paramletter}}
\newcommand{\jacmat}{\tns{J}}
\newcommand{\diag}[1]{\mathrm{diag}\left(#1\right)}
\newcommand{\lmres}{\tns{r}}
\newcommand{\findiffs}[2]{\dfrac{ \delta #1}{\delta #2}}

\newcommand{\mpindex}{\mathrm{mp}}

\newcommand{\graderr}{\mathrm{err}_\mathrm{grad}}
\newcommand{\reserr}{\mathrm{err}_\mathrm{res}}
\newcommand{\lmgradcrit}{\epsilon_{\mathrm{grad}}}
\newcommand{\lmrescrit}{\epsilon_{\mathrm{res}}}
\newcommand{\obsindex}{\mathrm{obs}}

\newcommand{\defgrad}{\tns{F}}
\newcommand{\defjac}{J}
\newcommand{\cauchy}{\tns{\sigma}}
\newcommand{\secpk}{\tns{{S}}}
\newcommand{\glstrain}{\tns{{E}}}
\newcommand{\eastrain}{\tns{\epsilon}}

\newcommand{\bodyforce}{\hat{\tns b}}
\newcommand{\strainenergy}{\psi}

\newcommand{\youngs}{\mathrm{E}}
\newcommand{\poissonratio}{\mathrm{\nu}}

\newcommand{\normal}{\tns{n}}
\newcommand{\tangent}{\tns{t}}

\newcommand{\domain}{\Omega}

\newcommand{\interface}{\Gamma}

\newcommand{\disp}{d}
\newcommand{\vel}{v}
\newcommand{\pres}{p}

\newcommand{\dens}{\rho}
\newcommand{\dynvis}{\mu}

\newcommand{\unit}[1]{\mathrm{#1}}
\newcommand{\unitfrac}[2]{\frac{\unit{#1}}{\unit{#2}}}

\newcommand{\sunit}[1]{\,\mathrm{#1}}
\newcommand{\sunitfrac}[2]{\,{\unit{#1}/\unit{#2}}}

\newcommand{\ddt}[1]{\dfrac{\mathrm{d} {#1}}{\mathrm{d} t}}
\newcommand{\ddtq}[1]{\dfrac{\mathrm{d}^2 {#1}}{\mathrm{d} t^2}}
\newcommand{\partialdt}[1]{\dfrac{ \partial #1}{\partial t}}
\newcommand{\partialdtdt}[1]{\dfrac{ \partial^2 #1}{\partial t^2}}

\newcommand{\structletter}{{\mathrm{S}}}

\newcommand{\disps}[1][]{\tns{\displacements}^{\structletter #1}}
\newcommand{\displacements}{\disp}

\newcommand{\fluidletter}{{\mathrm{F}}}

\newcommand{\velf}{\tns{\vel}^\fluidletter}
\newcommand{\presf}{\pres^\fluidletter}
\newcommand{\cauchyf}{\tns{\cauchy}^{\fluidletter}}
\newcommand{\densf}{\dens^{\fluidletter}}
\newcommand{\dynvisf}{\dynvis^{\fluidletter}}

\newcommand{\alevel}{\tns{c}^\fluidletter}

\newcommand{\conc}{\Phi}
\newcommand{\diffusivity}{D}
\newcommand{\react}{R}
\newcommand{\flux}{h}
\newcommand{\phasek}{\mathrm{K}}

\newcommand{\growthindex}{{\mathrm{g}}}

\newcommand{\poroletter}{{\mathrm{P}}}
\newcommand{\porofluidletter}{{\poroletter^\fluidletter}}
\newcommand{\porostructletter}{{\poroletter^\structletter}}

\newcommand{\porodisp}{\tns{\disp}^\porostructletter}
\newcommand{\poroxref}{\tns{X}^\poroletter}
\newcommand{\porovel}{\tns{\vel}^\porofluidletter}
\newcommand{\poropres}{\pres^\porofluidletter}

\newcommand{\fpipermeab}{\kappa}

\newcommand{\permeabpscalar}{k}
\newcommand{\poropermeab}{\tns{\permeabpscalar}}
\newcommand{\matpermeabp}{\tns{\matpermeabpscalar}}
\newcommand{\matpermeabpscalar}{K}

\newcommand{\porosity}{\phi}

\newcommand{\strainenergyp}{\psi^\poroletter}

\newcommand{\forwardmodelletter}{\mathfrak{M}}
\newcommand{\model}[1]{\forwardmodelletter\left(#1\right)}
\newcommand{\signeddist}{d}



%% file: table_initial_guesses.tex
\begin{tabular}{>{\centering\arraybackslash}p{1.5cm}>{\centering\arraybackslash}p{1.5cm}>{\centering\arraybackslash}p{1.5cm}}
\toprule
color & $\initialtop{\youngs} [\unit{Pa}]$ &   $\initialtop{\poissonratio} [-]$\\
\midrule
blue   &            200 &            0.4 \\
orange &            200 &            0.0 \\
green  &            400 &           0.45 \\
red    &            400 &            0.0 \\
purple &            600 &            0.4 \\
brown  &            600 &            0.0 \\
pink   &            400 &           -0.3 \\
gray   &            400 &           -0.6 \\
olive  &            400 &           -0.9 \\
cyan   &            1000 &           0.0 \\
black  &            1000 &           0.4 \\
\bottomrule
\end{tabular}

%% file: table_res_E_nue.tex
\begin{tabular}{>{\centering\arraybackslash}p{1.5cm}>{\centering\arraybackslash}p{1.8cm}>{\centering\arraybackslash}p{1.5cm}>{\centering\arraybackslash}p{1.5cm}>{\centering\arraybackslash}p{1.5cm}>{\centering\arraybackslash}p{1.5cm}>{\centering\arraybackslash}p{1.5cm}}
\toprule
\hspace{2mm}$\sigma$\newline$[\unit{mm}]$ & mean $\reserr$ \newline $[\unit{mm}]$ &  std $\reserr$ \newline $[\unit{mm}]$ &  mean $\youngs$\newline  $[\unit{Pa}]$ &  std $\youngs$ \newline $[\unit{Pa}]$ &  mean $\poissonratio$\newline  $[-]$ &  std $\poissonratio$ \newline $[-]$ \\
\midrule
                                0.000e+00 &                             7.122e-08 &                             3.120e-08 &                              4.000e+02 &                             1.226e-03 &                            3.000e-01 &                           6.889e-06 \\
                                1.000e-04 &                             1.478e-04 &                             6.576e-09 &                              3.999e+02 &                             8.615e-03 &                            3.015e-01 &                           6.881e-05 \\
                                1.000e-03 &                             9.296e-04 &                             2.686e-09 &                              4.080e+02 &                             1.168e-02 &                            3.782e-01 &                           1.153e-04 \\
                                1.000e-02 &                             8.208e-02 &                             5.892e-09 &                              5.062e+02 &                             5.044e-01 &                            2.998e-01 &                           3.591e-03 \\
\bottomrule
\end{tabular}

%% file: table_initial_guesses_hetero.tex
\begin{tabular}{>{\centering\arraybackslash}p{1.5cm}>{\centering\arraybackslash}p{1.5cm}>{\centering\arraybackslash}p{1.5cm}}
\toprule
color & $\initialtop{\youngs}_{\{1,2,3\}} [\unit{Pa}]$ &   $\initialtop{\poissonratio}_{\{1,2,3\}} [-]$\\
\midrule
blue   &            300 &            0.0 \\
orange &            500 &            0.3 \\
green  &            600 &            0.0 \\

\bottomrule
\end{tabular}

%% file: table_initial_guesses_hetero_homo.tex
\begin{tabular}{>{\centering\arraybackslash}p{1.0cm}>{\centering\arraybackslash}p{1.0cm}>{\centering\arraybackslash}p{1.0cm}}
\toprule
color & $\initialtop{\youngs} [\unit{Pa}]$ &   $\initialtop{\poissonratio} [-]$\\
\midrule
blue   &            200 &            0.0 \\
orange &            400 &            0.2 \\

\bottomrule
\end{tabular}

%% file: table_initial_guesses_poro.tex
\begin{tabular}{>{\centering\arraybackslash}p{1.5cm}>{\centering\arraybackslash}p{1.5cm}>{\centering\arraybackslash}p{1.5cm}}
\toprule
color & $\initialtop{\youngs} [\unit{Pa}]$ &   $\initialtop{\porosity} [-]$\\
\midrule
blue   &            400 &            0.2 \\
orange &            400 &            0.4 \\
green  &            300 &            0.6 \\

\bottomrule
\end{tabular}

%% file: table_initial_guesses_growth.tex
\begin{tabular}{>{\centering\arraybackslash}p{1.5cm}>{\centering\arraybackslash}p{1.9cm}>{\centering\arraybackslash}p{1.9cm}>{\centering\arraybackslash}p{1.9cm}}
\toprule
color & $\initialtop{\left(K_1^\growthindex\right)}\left[\unitfrac{mm^3}{mol}\right]$ &   $\initialtop{\left(K_2^\growthindex\right)}\left[\unitfrac{mm^2\,s}{g}\right]$ & $\initialtop{\left(K_3^\growthindex\right)}\left[\unitfrac{mm^2\,s}{g}\right]$\\
\midrule
blue   &            $10^{4}$ &            $10^{-2}$ &           $10^{-2}$ \\
orange &            $10^{5}$ &            $10^{-1}$ &           $10^{-1}$ \\
green  &            $10^{3}$ &            $10^{-3}$ &           $10^{-3}$ \\
red    &      $3\cdot10^{4}$ &      $8\cdot10^{-2}$ &     $5\cdot10^{-2}$ \\

\bottomrule
\end{tabular}